\documentclass{elsarticle}
\usepackage{amsmath,amssymb,a4wide,psfrag}
\newcommand{\Tr}[0]{\text{Tr}\;}
\newcommand{\sign}[0]{\text{sign}\;}

\renewcommand{\qed}{\nobreak \ifvmode \relax \else
      \ifdim\lastskip<1.5em \hskip-\lastskip
      \hskip1.5em plus0em minus0.5em \fi \nobreak
      \vrule height0.75em width0.5em depth0.25em\fi}

\begin{document}
\begin{frontmatter}
\title{A further extension to the group of Ginsparg-Wilson (overlap) chiral symmetries}
\author[a]{Nigel Cundy}
\author[a]{Weonjong Lee}
\address[a]{Lattice Gauge Theory Research Center, FPRD, and CTP,\\ Department of Physics \&
    Astronomy, Seoul National University, Seoul, 151-747,\\ South Korea}
\date{\today}
\begin{abstract}
Following work by Mandula, it has been observed that the Ginsparg Wilson lattice realisation of chiral symmetry has a possible problem: there is not just one lattice chiral symmetry, but an infinite group of lattice chiral symmetries with non-commuting generators. The physical implications of this abundance of symmetry remains unclear. In recent work, it has been shown how these chiral symmetries for overlap fermions can be derived from a Ginsparg-Wilson style renormalisation group blocking in the continuum, transforming the action from the standard continuum action to a lattice-like action similar to the action for overlap fermions. There is no unique blocking to obtain the overlap action from the continuum. Different blockings lead to different chiral symmetries; and the group of symmetries found by Mandula immediately follows. In this way, the excess chiral symmetry on the lattice can be explained in terms of different renormalisation schemes. 

The previous work suffered from one technical challenge: there is no continuum analogue of the lattice chiral eigenvectors at eigenvalue $2/a$, the zero mode doublers. Although it is not a necessary part of the formulation, to easily obtain standard results a simple blocking was chosen which mapped specific eigenfunctions of the continuum Dirac operator with eigenvectors of the lattice Dirac operator. For the zero modes of the Dirac operator and non-zero pairs, this poses no serious problems. But the zero mode doublers have no obvious counterpart in the continuum theory. Although the lattice chiral symmetry can still be defined, this leads to difficulties when considering $\mathcal{CP}$ symmetry on the lattice. In this work, we investigate the possibility of resolving this ambiguity by adding a second fermion field to the original continuum action used as a basis of the renormalisation group blockings. This second fermion field has a mass of the order of the momentum cut-off, to simulate the effects of the fermion doublers. Working through the same renormalisation group procedure to map this action to the lattice overlap action yields additional Ginsparg-Wilson relations satisfied by the overlap operator, and more (non-commuting) lattice chiral symmetries.
\end{abstract}
\begin{keyword}
Chiral fermions \sep Lattice QCD  \sep Renormalisation group
\PACS  11.30.Rd \sep 11.15.Ha \sep 11.10.Hi 
\end{keyword}
\end{frontmatter}
\section{Introduction}\label{sec:1}
Chiral symmetry in massless lattice QCD has been established through the Ginsparg-Wilson relation~\cite{Ginsparg:1982bj,Luscher:1998pqa} and practically implementable Dirac operators that satisfy this symmetry to any desired precision, the overlap~\cite{Narayanan:1993sk,Narayanan:1993ss,Neuberger:1998my,Neuberger:1998fp} and similar~\cite{Cundy:2008cs} Dirac operators. The presence (and spontaneous breaking) of chiral symmetry is the basis of the low energy effective theory that leads to chiral perturbation theory, as well as, on the lattice, restricting operator mixing and providing (for each overlap operator) a well defined topological framework.

However, there is an interesting and so far unresolved peculiarity affecting lattice chiral symmetry~\cite{Mandula:2007jt,Mandula:2009yd}: the overlap operator (and any other Ginsparg-Wilson Dirac operator, but in this work we shall just consider the overlap as a concrete example) obeys not just one but many lattice chiral symmetries. This can be seen relatively easily. The Ginsparg Wilson equation for a Dirac operator $D$ can be written as
\begin{gather}
\gamma_L^{(1)} D + D \gamma_R^{(1)} = 0,\label{eq:gw}
\end{gather}
where, in one well-used formulation, $\gamma_L^{(1)} = \gamma_5$, $\gamma_R^{(1)} = \gamma_5(1-D)$, the index $(1)$ refers to the particular choice of $\gamma_L$ and $\gamma_R$ for this chiral symmetry (using the notation of~\cite{Cundy:2010pu}), the massless overlap operator is
\begin{gather}
D = 1 + \gamma_5 \sign(K)
\end{gather}
and $K$ is a suitable doubler-free Hermitian Kernel operator, frequently (including in this work whenever a specific example is needed) taken to be a form of the Hermitian Wilson Dirac operator at a negative mass. As we are in Euclidean space-time, there is no difficulty in using different chiral rotations for the fermion and anti-fermion fields. Any $\gamma_L$ and $\gamma_R$, whether the conventional choice or another realisation of the symmetry, implies that the action is invariant under a chiral rotation
\begin{align}
\overline{\psi} \rightarrow& \overline{\psi}e^{i\eta \gamma_L}&{\psi} \rightarrow& e^{i\eta \gamma_R}{\psi}.
\end{align}
A topological index, conserved currents, and resolution of the U(1) anomaly all follow from this rotation from the usual methods.
There is a second well-known and well-used solution, $\gamma_L^{(-1)} = (1-D)\gamma_5$, $\gamma_R^{(-1)} = \gamma_5$, which is algebraically equivalent to equation (\ref{eq:gw}). 

Multiplying the Ginsparg-Wilson relation, equation (\ref{eq:gw}), from the left by $(\gamma_L^{(1)}\gamma_L^{(-1)})$ gives, after applying the Ginsparg-Wilson relation twice more,
\begin{gather}
\gamma_L^{(1)}\gamma_L^{(-1)}\gamma_L^{(1)} D + D \gamma_R^{(1)}\gamma_R^{(-1)}\gamma_R^{(1)} = 0.
\end{gather}
Since $[\gamma_L^{(1)},\gamma_L^{(-1)}]\neq 0$ and $[\gamma_R^{(1)},\gamma_R^{(-1)}]\neq 0$, this can be re-written as
\begin{gather}
\gamma_L^{(3)} D + D \gamma_R^{(3)} = 0,
\end{gather}
a third Ginsparg Wilson equation with a third corresponding lattice chiral symmetry. We can expand the group of Ginsparg-Wilson equations further:
\begin{align}
(\gamma_L^{(-1)}\gamma_L^{(1)})^n D - D (\gamma_R^{(-1)}\gamma_R^{(1)})^n =& 0,\nonumber\\
\gamma_L^{(1)}(\gamma_L^{(-1)}\gamma_L^{(1)})^n D + D \gamma_R^{(1)}(\gamma_R^{(-1)}\gamma_R^{(1)})^n =& 0,\label{eq:Mandulasymmetries}
\end{align}
and there is an infinite number of non-commuting lattice chiral symmetries, constructed with the operators
\begin{align}
\gamma_L^{(1+2n)} =& \gamma_L^{(1)} (\gamma_L^{(-1)} \gamma_L^{(1)})^n\nonumber\\
\gamma_R^{(1+2n)} =&  (\gamma_R^{(1)} \gamma_R^{(-1)})^n\gamma_R^{(1)}.
\end{align}
 This creates a number of conceptual problems, since there is also an infinite number of (presumably different) conserved currents, chiral Lagrangians and pions, one for each of these possible $\gamma_L$ and $\gamma_R$. All these chiral Lagrangians will reduce to the same continuum Lagrangian and one might expect that they are in some way equivalent on the lattice, but this has not yet been directly proved. 

In a series of papers, one of us has been working towards a different and seemingly unrelated goal of examining if there is an alternative derivation of the lattice overlap action using the same renormalisation group tools that are the basis of the Ginsparg-Wilson equation and the fixed point fermion. This goal still remains illusive, as the Yang-Mills part of the action or consider renormalisation group flows of the gauge field (both of which are related) have yet to be addressed. So far, only Wilson style renormalisation group blockings of the fermion fields have been considered. If the continuum generating functional (for one flavour of massless fermions) is
\begin{gather}
Z_0[J,\overline{J}] = \int dU \int d \psi_0 \int d\overline{\psi}_0 e^{-\frac{1}{4 g^2} F_{\mu\nu}^2 - \overline{\psi}_0 D_0 \psi_0 + \overline{\psi}_0J + \overline{J} \psi_0},
\end{gather} 
then introducing new fermion fields $\psi_1^{(\eta)}$ and a tunable real number $\Lambda$ using the ansatz
\begin{align}
Z_0[J,\overline{J}] = \int dU \int d \psi_0 \int d\overline{\psi}_0& e^{-\frac{1}{4 g^2} F_{\mu\nu}^2 - \overline{\psi}_0 D_0 \psi_0 + \overline{\psi}_0J + \overline{J} \psi_0}\frac{1}{\det\Lambda\alpha}\nonumber\\
& \int d\psi_1^{(\eta)} d\overline{\psi}_1^{(\eta)} e^{-(\overline{\psi}^{(\eta)}_1 - \overline{\psi}_0 (\overline{B}^{(\eta)})^{-1})\Lambda\alpha(({\psi}^{(\eta)}_1 - ({B}^{(\eta)})^{-1}{\psi}_0 )}.
\end{align}
$\eta$ is a real parameter which will later be used to distinguish different blockings. The blockings $B$ and $\overline{B}$ are functions of the gauge field and contain a Dirac structure.
If $\alpha$ is chosen to be the identity operator, and the limit $\Lambda\rightarrow\infty$ is taken, one can reconstruct a new generating functional
\begin{gather}
Z_1[J,\overline{J}] = \int dU \int d \psi_1 \int d\overline{\psi}_1 e^{-\frac{1}{4 (g')^2} F_{\mu\nu}^2 - \overline{\psi}_1 D \psi_1 + \overline{\psi}_1^{(\eta)} \overline{B}^{(\eta)}J + \overline{J} B^{(\eta)}\psi_1^{(\eta)}},
\end{gather}
where 
\begin{gather}
D = \overline{B}^{(\eta)} D_0 B^{(\eta)}
\end{gather}
I have assumed that it is possible to absorb the determinant ratio $\det (D_0/D)$ into a renormalisation of the gauge links and redefinition of the Yang-Mills term, which may lead of a modification of the Yang-Mills coupling from $g$ to $g'$ (if this assumption is invalid, the derivation and interpretation of the results of this work will break down, although the results themselves will stand on their own. It is beyond the scope of this work to investigate whether this assumption is correct for the actions under consideration; we are only investigating some consequences should it prove to hold. We can expect it to be valid if $\Tr \log (D/D_0)$ is local, i.e. the Fourier transform of $\Tr \log (D/D_0)$ remains analytic). 

This procedure can be used to construct a new, exponentially local, Dirac operator $D$, which may have a very different functional form to the standard continuum Dirac operator. It is, of course, impossible to construct a reversible blocking from the continuum to the lattice (i.e. where $B$ and $\overline{B}$ are invertible so $D_0 = \overline{B}^{-1} D B^{-1}$), since the lattice involves the loss of a number of degrees of freedom; however it is possible to construct a continuum Dirac operator which resembles the lattice overlap operator in that it has the same dispersion relation. The basic idea~\cite{Cundy:2009ab} is to give the extra degrees of freedom a mass of the order of the cut-off, similar to the method used by Wilson to removed the extra degrees of freedom caused by the fermion doublers. A continuum  Wilson Dirac operator is constructed by blocking from the continuum, regulated to avoid unwanted poles away from the lattice sites (so that the Dirac operator e.g. decays exponentially around the lattice sites rather than is a sum over delta functions on the lattice sites as in a pure lattice theory), lattice modes are decoupled from off-lattice eigenvectors which are given an infinite mass, and placed inside the overlap formula, which has the effect of eliminating the additive mass renormalisation coming from the treatment of the off-lattice eigenvectors. Finally, a smooth limit is taken to recover the lattice theory. The blockings must therefore be functions of the gauge field and, to avoid massless fermion doublers, contain some Dirac structure.   

Once an equivalent to the lattice overlap operator has been constructed in the continuum in this way, finding blockings to create the correct action is straight-forward: one can use, for example
\begin{align}
B^{(1)} =& D_0^{-1} Z D\nonumber\\
\overline{B}^{(1)} = & Z^{\dagger},\label{eq:firstblockings}
\end{align}
where $Z$ is a unitary operator (arbitrary within the constraint that the blocking and its inverse must exist and remain local) which can be used to map in some way between the standard continuum operator and the continuum equivalent to the lattice Dirac operator. Clearly these blockings will give $\overline{B}^{(1)}D_0B^{(1)} = D$, and the target lattice action (neglecting again the effect of integrating out the fermion fields). In the previous work,  a spectral mapping between the two Dirac operators was used, which greatly simplifies the calculation of the chiral symmetry operators.

Following the Ginsparg Wilson procedure~\cite{Ginsparg:1982bj,Cundy:2009ab}, it can be shown that this blocking implies a lattice chiral symmetry, with 
\begin{align}
\gamma_L^{(\eta)} =& \overline{B}^{(\eta)} \gamma_5 (\overline{B}^{(\eta)})^{-1}\nonumber\\
\gamma_R^{(\eta)} =& ({B}^{(\eta)})^{-1} \gamma_5 {B}^{(\eta)}.
\end{align} 
With $Z$ chosen according to~\cite{Cundy:2009ab}, the blockings in equation (\ref{eq:firstblockings}) give $\gamma_L^{(1)} = \gamma_5$ and $\gamma_R^{(1)} = \gamma_5(1-D)$, the standard Ginsparg-Wilson symmetry. However, these blockings are not unique, and one can equally choose
\begin{align}
B^{(\eta)} =& D_0^{-(1+\eta)/2} Z D^{(1+\eta)/2}\nonumber\\
\overline{B}^{(\eta)} = & D^{(1-\eta)/2}Z^{\dagger}D_0^{-(1-\eta)/2},\label{eq:secondblockings}
\end{align}
for any real $\eta$
and these blockings will give a different lattice chiral symmetry, implemented through the operators~\cite{Cundy:2010pu}
\begin{align}
{\gamma}^{(\eta)}_R = & \gamma_5 \cos\left[\frac{1}{2}(1+\eta)(\pi - 2 \theta)\right] + \sign(\gamma_5 (D^{\dagger}-D)) \sin\left[ \frac{1}{2}(1+\eta)(\pi - 2 \theta)\right]\nonumber\\
{\gamma}^{(\eta)}_L = & \gamma_5 \cos\left[\frac{1}{2}(\eta-1)(\pi - 2 \theta)\right] + \sign(\gamma_5(D^{\dagger}-D)) \sin\left[ \frac{1}{2}(\eta-1)(\pi - 2 \theta)\right]\label{eq:8a}\\
\tan\theta =& 2 \frac{\sqrt{1-a^2D^{\dagger} D/4}}{\sqrt{a^2 D^{\dagger} D}}.\label{eq:gamma_eta0}
\end{align}
Both these lattice $\gamma_5$ operators and the blockings are only local and well defined when $\eta$ is an odd integer. It can be shown that
\begin{gather}
\gamma_R^{(\eta_1)}\gamma_R^{(\eta_2)} = \gamma_5 \gamma_R^{(\eta_2-\eta_1 -1)},
\end{gather} 
and $\gamma_R^{(\eta)}\gamma_R^{(\eta)} = 1$ thus
\begin{gather}
\gamma_R^{(\eta + 2)} = \gamma_R^{(\eta )}\gamma_5 \gamma_R^{(1)},
\end{gather}
and it is clear that these chiral symmetries for odd integer $\eta$ are identical to the Mandula chiral symmetries of equation (\ref{eq:Mandulasymmetries}).

While this approach to understanding the origin of the excess chiral symmetry on the lattice seems natural, there are  a few related anomalies in the construction. The most obvious was encountered in~\cite{Cundy:2010pu}, which attempted to incorporate a lattice $\mathcal{CP}$ symmetry~\cite{Igarashi:2009wr} into this framework: the lattice $\mathcal{CP}$ operators are non-local in the presence of the zero mode doublers (the eigenvalues at 2 of the overlap operator), although the $\mathcal{CP}$ symmetry transformations, action and all operators describing observables in the chiral gauge theory and standard theory are themselves local and well defined. The result of this non-locality concerns the continuum limit of the zero mode doublers $\phi_2$ (which are the eigenvectors which satisfy $D\phi_2 = 2 \phi_2$): the continuum limit of the $\mathcal{CP}$ transformation of  $\gamma_R^{(\eta)}\phi_2 \phi_2^{\dagger}$ does not reduce to the $\mathcal{CP}$ transformation of $\gamma_5\phi'_2 {\phi'_2}^{\dagger}$, where $\phi'_2$ represent some continuum equivalent of the zero mode doublers because there is no continuum equivalent of the zero mode doublers. In the continuum theory, the only eigenvectors of both $\gamma_5$ and the Dirac operator are the zero modes themselves. It is therefore not surprising that some anomalies should appear while taking the continuum limit of $\gamma_R^{(\eta)}\phi_2 \phi_2^{\dagger}$. In the initial work, there was an implicitly assumption that the zero mode doublers could be paired with a linear combination of the infinite number of continuum modes not seen on the lattice; for the construction of the chiral symmetry operators this was good enough. However, once we started to consider the case of $\mathcal{CP}$ on the lattice, it became clear that a different approach would be better. Additionally, as shall be mentioned below, the continuum Dirac operator only represents the first half of the lattice overlap eigenvalue circle.

One obvious solution to these problems is to construct a continuum action with doublers, and apply the renormalisation group blockings to that action to obtain the lattice theory. This is permissible: the mass of these doublers will be of the order of the momentum cut-off used to regulate QCD, and therefore their presence will not affect any experimental observable. Usually this action would be buried within the higher order irrelevant terms of a Symanzik expansion of the lattice QCD action: one of many cut off effects that are usually (and rightly) neglected. Here, however, we include it.

In this work, we explore whether it is possible to map from such a continuum theory to the lattice via the renormalisation group. If successful, one consequence of this may be a new group of lattice chiral symmetries. In section \ref{sec:2}, we consider the term that needs to be added to the continuum action, and consider the chiral symmetries and also $\mathcal{CP}$ symmetry and the chiral gauge theory, as these were where the anomalies in the original construction first appeared. Then, in \ref{sec:3}, we construct an continuum overlap action with the standard continuum Dirac operators as its kernels. In \ref{sec:4}, we extend this to the continuum equivalent of lattice overlap fermions. After concluding, there is one appendix which discusses the locality of the new lattice $\gamma_5$ and $\mathcal{CP}$ operators.

\section{Eigenvalue decomposition}\label{sec:2}
For all the Dirac operators considered in this work, the Hermitian square of the operator commutes with $\gamma_5$. This means that those eigenvectors of the Dirac operator which do not commute with $\gamma_5$ (which is all of them except the exact zero modes and their doublers at eigenvalue 2) are paired, with each pair constructed from the same two degenerate chiral eigenvectors of $D^{\dagger}D$. We use the chiral representation for $\gamma_5$
\begin{gather}
\gamma_5 = \left(\begin{array}{l l}\phi_+ & \phi_-\end{array}\right)\left(\begin{array}{l l} 1&0\\0&-1\end{array}\right)\left(\begin{array}{l}\phi_+^{\dagger} \\ \phi_-^{\dagger}\end{array}\right)\nonumber.
\end{gather} 
For the moment, $D$ represents any Dirac operator with a chiral symmetry.  The eigenvalues of $D^{\dagger}D$ are $\lambda^2_i$ with chiral eigenvectors $\phi_{\pm i}$, since $[\gamma_5,D^{\dagger}D] = 0$ (which can be proved from the Ginsparg-Wilson for any possible $\gamma_L$, and for each chiral Dirac operator $\gamma_5$ is a possible choice for $\gamma_L$). $D$ can be written in a basis constructed from the eigenvector pairs
\begin{gather}
D^{\dagger} D = \left(\begin{array}{l l}\phi_+ & \phi_-\end{array}\right)\left(\begin{array}{l l} \lambda^2&0\\0&\lambda^2\end{array}\right)\left(\begin{array}{l}\phi_+^{\dagger} \\ \phi_-^{\dagger}\end{array}\right).
\end{gather}
For most of this work, we shall adopt a notation which omits the chiral eigenvectors $\phi_+$ and $\phi_-$ in this matrix representation of the Dirac operators.

If we insist that $D$ is $\gamma_5$-Hermitian, then (up to a phase on the off diagonal terms which can be absorbed into the eigenvectors) in this spectral basis
\begin{gather}
D = \lambda \left(\begin{array}{l l} \cos\theta&\sin\theta\\-\sin\theta&\cos\theta\end{array}\right)
\end{gather}
We can also define $\phi_{0i}$ as the zero modes of $D$ with eigenvalue $\varepsilon = 0$ and $\phi_{2i}$ as their doublers with eigenvalue $\Lambda$ (for those Dirac operators which have doublers). This matrix representation of the massless Dirac operator is equivalent to the spectral decomposition
\begin{gather}
D = \sum_i \varepsilon \phi_{0i} \phi_{0i} + \sum_i \Lambda \phi_{2i}\phi_{2i} + \sum_i \lambda_i\left[\phi_{+i}\phi_{+i}^{\dagger} \cos{\theta}_i + \phi_{-i}\phi_{-i}^{\dagger} \cos{\theta}_i+\phi_{+i}\phi_{-i}^{\dagger} \sin{\theta}_i - \phi_{-i}\phi_{+i}^{\dagger} \sin{\theta}_i\right].
\end{gather}
This notation can be extended to cover the massive Dirac operator (the zero modes are explicitly included in this definition of $D$ to facilitate the adaptation to the massive case where $\varepsilon \neq 0$).
For the standard continuum Dirac operator, which is anti-Hermitian, $\theta = \pi/2$. For the lattice overlap Dirac operator, $\cos\theta = \lambda/2$.

The spectral decomposition is, of course, impossible to use in practice. However the matrix representation also allows for an easy conversion to a more practical expression for the Dirac operator, since, for the non zero modes (by which we mean those eigenvectors whose eigenvalues for the massless operator are not equal to zero or two)
\begin{align}
\left(\begin{array}{l l} 1&0\\0&1\end{array}\right) =& 1,&
\left(\begin{array}{l l} 1&0\\0&-1\end{array}\right) =& \gamma_5,\nonumber\\
\left(\begin{array}{l l} 0&1\\1&0\end{array}\right) =& \sign(\gamma_5(D-D^{\dagger})),&
\left(\begin{array}{l l} 0&1\\-1&0\end{array}\right) =& \gamma_5\sign(\gamma_5(D-D^{\dagger})).
\end{align} 
One at most needs to treat the zero modes and their doublers via deflation; for a local operator this will not be required since the functional form extracted for the non-zero modes will also be valid for the zero modes and their doublers.

\section{The Continuum action}\label{sec:3}

In the previous work, Wilson's blocking formulation of the renormalisation group was used to map between the overlap lattice action and a single flavour of continuum fermions. However, the lattice action describes two fermion fields, the physical field and a doubler field with an mass of the order of the cut-off, while the standard continuum action just describes one. Although this subtlety proved unimportant when constructing the chiral symmetries of the theory~\cite{Cundy:2009ab}, the fact that the lattice zero mode doubler lacked a continuum counterpart was discomforting. The formulation worked for the zero mode doubler, but without a sound theoretical reason for expecting that it should work. When the work was subsequently extended to include $\mathcal{CP}$ symmetry~\cite{Cundy:2010pu}, this  proved to be a more serious problem: not that the lattice $\mathcal{CP}$ generators was non-local (which itself is no surprise since even the operators which generate continuum $\mathcal{CP}$ symmetry are non-local), but that the lattice $\mathcal{CP}$ did not reduce to continuum $\mathcal{CP}$ for the eigenvalues at $D=2$, the zero mode doublers, as demonstrated in a non-locality in the operator mapping lattice $\mathcal{CP}$ to continuum $\mathcal{CP}$ when the doublers are present. This should be expected: one cannot take the continuum limit of a lattice $\mathcal{CP}$ transformation for doublers because there are no doublers in the continuum theory. 

Once this root of the problem is understood, the obvious solution is to add a doubler quark field to the continuum action. If the mass of this doubler field is of the order of the momentum cut-off, then the presence of this field will leave all observables unchanged, and we can do so with impunity since all experimental results will be unaffected. The new continuum fermion action reads
\begin{gather}
\overline{\psi}_0 D_0 \psi_0 + \overline{\psi}_d (D_d + 2/a) \psi_d,
\end{gather}
where $D_d$ is the massless Dirac operator with a $\mathcal{CP}$ transformed gauge field,
\begin{gather}
D_d\psi = (U^{CP})^{\dagger} \gamma_{\mu} \partial_{\mu}(U^{CP}\psi).
\end{gather}
$a$ will, in due course, become the lattice spacing, but for now remains only as  the inverse of a momentum cut-off (an ultra-violet regulator). The eigenvalues of $D_0$ and $D_d$ are thus in the range $0 < |i\lambda_0|<2/a$ and $0 < |i\lambda_d|<2/a$.
Using the $\mathcal{CP}$ transformed gauge field ensures that the zero modes of $D_d$ are in the opposite chiral sector to the zero modes of $D_0$ and thus have the same chirality as the zero mode doublers of the lattice overlap operator. In matrix form, we decompose $D_0$ as
\begin{gather}
D_0 = \lambda_0 R(\pi/2),
\end{gather}
and $D_d + 2$ as
\begin{gather}
a(D_d + 2) = \frac{2}{\cos\alpha} R(\alpha) 
\end{gather}
with
\begin{align}
\sin \alpha =& \frac{a\lambda_d}{\sqrt{4+a^2\lambda_d^2}}& \cos\alpha =& \frac{2}{\sqrt{4+a^2\lambda_d^2}}.
\end{align}
where we have defined
\begin{gather}
R(\alpha) = \left(\begin{array}{l l} \cos{\alpha}&\sin{\alpha}\\-\sin\alpha&\cos\alpha\end{array}\right).
\end{gather}
For simplicity, we will often subsequently use units where $a=1$. 

We can also consider a rotation of the physical continuum fields using a blocking defined in terms of a function $\alpha'$ of $D_0^{\dagger} D_0$.
\begin{align}
\psi'_0 =& R((\alpha'-\pi/2)/2) \psi_0\nonumber\\
\overline{\psi}'_0 = &\overline{\psi}_0 R(-(\alpha'-\pi/2)/2),  
\end{align}
where $\alpha' = \pi/2$ at the zero modes of the Dirac operator. As long as $\alpha'$ is chosen so that $R((\alpha'-\pi/2)/2)$ and $R(-(\alpha'-\pi/2)/2)$ are local, this rotation leaves the continuum action invariant, but will modify the $\mathcal{CP}$ symmetry and chiral symmetry.

\subsection{Chiral symmetry}\label{sec:3.1}
The action is invariant under the following infinitesimal chiral symmetry
\begin{align}
\psi_0 \rightarrow&\psi_0 + i \epsilon \gamma_5 \psi_0 & \overline{\psi}_0 \rightarrow&\overline{\psi}_0 + i \epsilon \overline{\psi}_0\gamma_5 \nonumber\\
 \psi_d \rightarrow&\psi_d + i \epsilon \gamma_{Rd} \psi_d & \overline{\psi}_d \rightarrow&\overline{\psi}_d + i \epsilon \overline{\psi}_d\gamma_{Ld}, 
\end{align}
with (for example)
\begin{align}
\gamma_{Rd} = &\left(1-\frac{4}{D_d + 2}\right)\gamma_5\nonumber\\
\gamma_{Ld} = &\gamma_5\label{eq:30}.
\end{align}
This formulation of chiral symmetry is not unique; as one example one can use instead
\begin{align}
\tilde{\gamma}_{Ld} = &\gamma_5\left(1-\frac{4}{D_d + 2}\right)\nonumber\\
\tilde{\gamma}_{Rd} = &\gamma_5,
\end{align}
or an infinite number of alternatives constructed following the argument outlined earlier. However, in this work, we shall only consider the symmetry outlined in equation (\ref{eq:30}).

In matrix form, we can write that 
\begin{gather}
\gamma_{Rd} = -\gamma_5 R(2\alpha),
\end{gather}
and it is easy to see that $\gamma_{Rd}^2 = \gamma_{Ld}^2 = 1$. Equally, in Euclidean space, $\gamma_{Rd}$ and $\gamma_{Ld}$ are both local.

For the modified action $\overline{\psi}'_0 D_0 \psi'_0$ the $\gamma$ matrices are $\gamma_{L0}' =  R(\pi/2-\alpha')\gamma_5$ and $\gamma_{R0}' = \gamma_5 R(\alpha' - \pi/2)$
\subsection{$\mathcal{CP}$ symmetry}\label{sec:3.2}
The action is invariant under the $\mathcal{CP}$ symmetry 
\begin{align}
D_0[U,x,y] \rightarrow& W D_0[U^{CP},\bar{x},\bar{y}]^T W^{-1}&\gamma_5 \rightarrow& -W \gamma_5^TW^{-1}\nonumber\\
\psi_0(x) \rightarrow &-W^{-1} (\overline{\psi}_0(\overline{x}))^T & \overline{\psi}_0(x) \rightarrow & ({\psi_0}(\overline{x}))^TW\nonumber\\ 
D_d[U^{CP},x,y] \rightarrow& W D_d[U,\bar{x},\bar{y}]^T W^{-1}&\gamma_5 \rightarrow& -W \gamma_5^TW^{-1}\nonumber\\
\psi_d(x) \rightarrow &- W^{-1}(\overline{\psi}_d(\overline{x})\Gamma_d^{\dagger})^T & \overline{\psi}_d(x) \rightarrow & (\Gamma_d{\psi}_d(\overline{x}))^TW,\label{eq:cpforpsid}
\end{align}
where $T$ denotes the transpose and
\begin{align}
\Gamma_d =& (\cos(\pi/2 - \alpha) - \gamma_5 \sin(\pi/2-\alpha) \sign(\gamma_5(D_d - D_d^{\dagger})))(1-\phi_{2d}\phi_{2d}^{\dagger}) + \phi_{2d} \phi_{2d}^{\dagger}\nonumber\\
=&R(\alpha-\pi/2) +\phi_{2d} \phi_{2d}^{\dagger}\label{eq:Gammacont1}.
\end{align}
The operator $W$ 
$\phi_{2d}$ are the zero eigenvectors of $D_d$. It will be observed that $\Gamma_d$ is not local; however, since this operator does not appear in any observables or actions (and, of course, the Parity operation itself is non-local) this will not affect any physical observable and is thus unimportant. We have no reason to require or expect that the operators defining $\mathcal{CP}$ symmetry should be local. Note that
\begin{align}
[\Gamma_d,D_d] = & [\Gamma_d^{\dagger},D_d] =0\nonumber\\
\Gamma_d^{\dagger} \gamma_5\Gamma_d =& \gamma_{Rd}\nonumber\\
\Gamma_d^{\dagger}\Gamma_d = &1,\label{eq:15}
\end{align}
and therefore the action is invariant under this $\mathcal{CP}$ symmetry.

For the modified action $\overline{\psi}'_0 D_0 \psi'_0$, the $\mathcal{CP}$ symmetry is defined in terms of 
\begin{gather}
\Gamma_0 = R(\alpha' - \pi/2).\label{eq:Gammacont2}
\end{gather}
\subsection{Chiral gauge theories}\label{sec:3.3}
The chiral gauge theory for right handed fermions and doublers can be written as
\begin{gather}
L = \frac{1}{2}\overline{\psi}_0 D_0 (1+\gamma_5) \psi_0 + \frac{1}{2}\overline{\psi}_d (D_d+2) (1+\gamma_{Rd}) \psi_d.
\end{gather} 
Using equations (\ref{eq:cpforpsid}) and (\ref{eq:15}), it is clear that the chiral gauge theory action is also invariant under both chiral and $\mathcal{CP}$ symmetries.

The measure of the doubler weyl fermion should be gauge invariant and invariant under $\mathcal{CP}$. This is not immediately obvious, since $\gamma_{Rd}$ is a function of the gauge fields. Following~\cite{Cundy:2010pu}, we can construct the measure in terms of the eigenvectors $g^{\pm}$ of $\gamma_{Rd}$ and $H_{\pm}$ of $\gamma_5 D$, 
\begin{align}
\gamma_{Rd} g^{\pm} =& \pm g^{\pm}\nonumber\\
\gamma_5 D_d H_{\pm} = & \pm \lambda H_{\pm}\nonumber\\
D_d^{\dagger}D_dg^{\pm} =& \lambda^2 g^{\pm}, 
\end{align}
where we have suppressed the eigenvalue index.

We can write the measure for the right handed fermion as $\prod_i c_i$, where $c_i$ are the coefficients of the right handed fermion field in the basis defined by $g^{+}$ 
\begin{gather}
\frac{1}{2}(1+\gamma_{Rd})\psi_d = \sum_i c_i g^{+}_{i} + \sum_i c_{0i} g_{0i},\label{eq:aaa1}
\end{gather}
where $g_{0i}$ are the zero modes of $D_d$ in the appropriate chiral sector.
The eigenvectors of $\gamma_{Rd}$ can be expressed in terms of the eigenvectors of $\gamma_5 D_d$ as
\begin{gather}
g^+ = \cos(\alpha/2) H_+ + \sin(\alpha/2) H_-\nonumber\\
g^- = \cos(\alpha/2) H_- - \sin(\alpha/2) H_+
\end{gather}
The $\mathcal{CP}$ transformations of these vectors can be calculated from the eigenvalue equations
\begin{align}
H_+ \rightarrow& W(H_-^{\dagger} \gamma_5)^T  & H_-\rightarrow& -W(H_+^{\dagger} \gamma_5)^T\nonumber\\
g^+ \rightarrow& W((g^-)^{\dagger} \gamma_5)^T  & g_-\rightarrow& -W((g^+)^{\dagger} \gamma_5)^T.
\end{align}
Using $\Gamma_d^{\dagger} \gamma_5\gamma_{Ld} \gamma_5\Gamma_d = \gamma_{Rd}$, a suitable measure for the left handed anti-fermion field is $\prod d\overline{c}_i$, where $\overline{c}$ are the appropriate analogues of $c$:  
\begin{gather}
\frac{1}{2}\overline{\psi}_d (1-\gamma_{Ld}) = \sum_i \overline{c}_i (g^{-}_{i})^{\dagger} \Gamma_d^{\dagger}\gamma_5 + \sum_i \overline{c}_{0i} g_{0i}^{\dagger}.\label{eq:aaa2}
\end{gather} 
Under $\mathcal{CP}$, $\Gamma_d$ transforms to $(\Gamma_d^{\dagger})^{T}$ and equations (\ref{eq:aaa1}) and (\ref{eq:aaa2}) become
\begin{gather}
\frac{1}{2}\overline{\psi}_d(1-\gamma_5)\Gamma_d^{\dagger} = -\sum_i c^{CP}_i (g^{-}_{i})^{\dagger}\gamma_5\nonumber\\
\frac {1}{2} \Gamma_d (1+\gamma_{Rd}) \psi_d = -\sum_i \overline{c}^{CP}_i \gamma_5\Gamma_d \gamma_5 g^+_{i}.
\end{gather}
Thus $c^{CP}_i = -\overline{c}_i$ and $\overline{c}^{CP}_i = -c_i$ and the measure for the $\mathcal{CP}$ transformation of the Weyl fields is unchanged.

Gauge invariance of the measure can be proved by considering how the eigenvalues transform under infinitesimal transformations of the fermion field~\cite{Cundy:2010pu}. The changes to the eigenvectors $H_+$ and $H_-$ after an infinitesimal change in the gauge field which induces a change $\delta D_d$ in the Dirac operator and $\delta \alpha$ in $\alpha$ are 
\begin{align}
\delta H_+ =& \frac{1}{\gamma_5 D_d - \lambda} (1-H_+ H_+^{\dagger}) \gamma_5 \delta D_d H_+\nonumber\\
\delta H_- =& \frac{1}{\gamma_5 D_d + \lambda} (1-H_- H_-^{\dagger}) \gamma_5 \delta D_d H_-.\label{eq:evdiff}
\end{align}
Thus,
\begin{align}
\delta g^+  =& \frac{\delta \alpha}{2} (\cos(\alpha/2) H_- - \sin(\alpha/2) H_+) + \nonumber\\
&\frac{\cos(\alpha/2)}{\gamma_5 D_d - \lambda}(1-H_+ H_+^{\dagger}) \gamma_5 \delta D_d H_+ + \frac{\sin(\alpha/2)}{\gamma_5 D_d + \lambda}(1-H_- H_-^{\dagger}) \gamma_5 \delta D_d H_-.
\end{align}
The change in the measure induced by this change in the basis is~\cite{Luscher:1998du} $\mathcal{L} = \sum_i (g^+_{i},\delta g^+_{i})$. The contribution to $\mathcal{L}$ from the zero modes and their doublers is zero, as $(g_0,\delta g_0) = (\phi_2,\delta \phi_2) = 0$ (for one zero mode, this follows using the same eigenvector differentiation employed in equation (\ref{eq:evdiff}); for degenerate zero modes a more careful differentiation of the eigenvectors needs to be employed, extending the methods of ~\cite{Cundy:2007df}, but the same result holds). For the non-zero eigenvalues (where the measure for degenerate eigenvectors of $\gamma_5 D_d$ is zero)
\begin{align}
\mathcal{L} =& \sum_i \frac{\delta \alpha_i}{2}(\cos(\alpha_i/2) \sin(\alpha_i/2) - \sin(\alpha_i/2) \cos(\alpha_i/2)) + \frac{\sin(\alpha_i)}{4\lambda}(H_{+i}^{\dagger} \gamma_5 \delta D_d H_{-i} - H_{-i}^{\dagger} \gamma_5 \delta D_d H_{+i}^{\dagger})\nonumber\\
=& \Tr\left(\gamma_5 \delta D_d \sum_i \frac{\sin(\alpha_i)}{4\lambda} (H_{-i}H_{+i}^{\dagger} - H_{+i}H_{-i}^{\dagger})\right)\nonumber\\
=& -\frac{1}{2}\Tr\left[\gamma_5 \delta D_d (D_d-D_d^{\dagger})\frac{1}{D^{\dagger} D}\right],
\end{align}
since
\begin{gather}
D_d^{\dagger}-D_d = 2 \lambda \sin\alpha (H_{-i}H_{+i}^{\dagger} - H_{+i}H_{-i}^{\dagger}). 
\end{gather}
Under an infinitesimal gauge transformation,
\begin{gather}
A_{\mu} \rightarrow A_{\mu} - \nabla_{\mu} \omega(x),
\end{gather}
we can write that~\cite{Luscher:1998du}
\begin{gather}
\delta D_d = [R_{\omega},D],
\end{gather}
where $R_{\omega}$ is the representation of the Lie algebra, and
\begin{gather}
\mathcal{L} = -\frac{1}{2} \Tr_{non-zero} \gamma_5 R_{\omega} (D_d-D_d^{\dagger})^2\frac{1}{D^{\dagger} D} = 0,
\end{gather}
since $D^{\dagger} D$ and $(D_d-D_d^{\dagger})^2$ both commute with $\gamma_5$.

The anomaly can be constructed in the usual way by considering the change in the measure under a chiral rotation.
\section{Continuum overlap fermions}\label{sec:4}
As an intermediate step between the standard continuum operator and the lattice, we now choose to use a renormalisation group blocking to introduce a new pair of Dirac operators in the continuum, $D_1$ and $D_2$.
\begin{align}
a D_1 =& 1 + \gamma_5 \sign \gamma_5 \left(\frac{a^2D_0}{1-a^2D_0^{\dagger} D_0/4} - m\right)\nonumber\\
a D_2 = & 1-\gamma_5 \sign \gamma_5 \left(\frac{a^2D_d^{\dagger}}{1-a^2D_d^{\dagger}D_d/4} - m\right),
\end{align} 
where $m$ is a tunable constant in the range $0<m<2$. $m>0$ is required to ensure that the pole in the propagator is at the correct momentum (i.e. at the same momentum as the pole in the propagator of the continuum fermion), and $m<2$ is required to ensure that the $\gamma_L$ and $\gamma_R$ operators are local. We shall assume (leaving the important discussion until a future work)  that a similar analysis to that of~\cite{Cundy:2009ab} will hold, and that the blockings are local and invertible, with any determinant from the blockings being absorbed into a renormalisation of the gauge fields and their couplings.\footnote{In~\cite{Cundy:2009ab}, it was demonstrated for the conversion to the lattice overlap, the locality of the blockings required that both Dirac operators had the same low momentum behaviour and no doublers, which is the case here, and there is no reason to suppose that a similar argument will not hold in this case. The issue of whether the determinant can be absorbed into a renormalisation of the gauge fields is more difficult, and would require an unnecessary lengthy digression from the focus of this work. It seems likely in this case because $\log(D_1/D_0)$ and $\log(D_2/(D_d+2))$ are both local, which means that the trace of these operators reduces to the Yang Mills action plus some irrelevant terms. Until this proof is found, the RG derivation and interpretation of the chiral symmetry and $\mathcal{CP}$ transformations must be considered tentative, although the final results stand on their own without this derivation.} The matrix representation of these Dirac operators is
\begin{align}
D_1 =& 2 \cos (\phi_0) R(\phi_0)\\
D_2 =& 2 \cos (\phi_d) R(\phi_d)
\end{align}
where
\begin{align}
\sin(2\phi_0) =& \frac{2\lambda_0}{\sqrt{(1-\lambda_0^2/4)^2m^2 + 4\lambda_0^2}}& \cos(2\phi_0) =&- \frac{m(1-\lambda_0^2/4)}{\sqrt{(1-\lambda_0^2/4)^2m^2 + 4\lambda_0^2}}\nonumber\\
\sin(2\phi_d) =& \frac{2\lambda_d}{\sqrt{(1-\lambda_d^2/4)^2m^2 + 4\lambda_d^2}}& \cos(2\phi_d) =& \frac{m(1-\lambda_d^2/4)}{\sqrt{(1-\lambda_d^2/4)^2m^2 + 4\lambda_d^2}}
\end{align}
For small $\lambda_0$ and $\lambda_d$, to terms up to O($\lambda$) and a multiplicative constant, $D_1$ and $D_2$ are the same as $D_0$ and $D_d$. $\phi_d$ is in the range $0\le\phi_d\le\pi/4$ while $\phi_0$ is in the range $\pi/4 \le \phi_0 \le \pi/2$. 

 One set of blockings which will generate the target action is 
\begin{align}
\tilde{B}_1^{(\eta)} =& D_0^{-(1+\eta)/2} D_1^{(1+\eta)/2}\nonumber\\
\tilde{\overline{B}}_1^{(\eta)} =&  D_1^{(1-\eta)/2}D_0^{-(1-\eta)/2}\nonumber\\
\hat{B}_2^{(\eta)} =& (2+D_d)^{-(1+\eta)/2} D_2^{(1+\eta)/2}\nonumber\\
\hat{\overline{B}}_2^{(\eta)} =&  D_2^{(1-\eta)/2}(D_d+2)^{-(1-\eta)/2},
\end{align}  
for real $\eta$,
which gives
\begin{align}
\tilde{B}_1^{(\eta)} =& \left(\frac{2 \cos\phi_0}{\lambda_0} \right)^{(1+\eta)/2}R((1+\eta)/2(\phi_0 - \pi/2))\nonumber\\
\tilde{\overline{B}}_1^{(\eta)} =& \left(\frac{2 \cos\phi_0}{\lambda_0} \right)^{(1-\eta)/2}R((1-\eta)/2(\phi_0 - \pi/2))\nonumber\\
\hat{B}_2^{(\eta)} =& \left(\cos\phi_d \cos\alpha\right)^{(1+\eta)/2}R((1+\eta)/2(\phi_d - \alpha))\nonumber\\
\hat{\overline{B}}_2^{(\eta)} =& \left(\cos\phi_d \cos\alpha \right)^{(1-\eta)/2}R((1-\eta)/2(\phi_d - \alpha)).\label{eq:naive}
\end{align}
$B_1$ and $\overline{B}_1$ are the blockings used to generate the group of lattice chiral symmetries described  in~\cite{Cundy:2010pu}. This $B_2$ and $\overline{B}_2$ seem a natural extension to the larger picture described here; however, since $D_d +2$ does not have the standard chiral symmetry, this choice of blockings would lead to discontinuous and non-local $\gamma_R$ and $\gamma_L$ matrices. Better (for those values of $\eta$ where the these blockings remain local) is to use the choice
\begin{align}
\tilde{B}_2^{(\eta)} =&\left(\cos\phi_d \cos\alpha \right)^{(1+\eta)/2} R( (1+\eta)(\phi_d - \alpha')/2 -(1-\eta)(\alpha' - \pi/2)/2)\nonumber\\
\tilde{\overline{B}}_2^{(\eta)} =& \left(\cos\phi_d \cos\alpha \right)^{(1-\eta)/2}R((1-\eta)(\phi_d - \alpha')/2 + (1-\eta)(\alpha' - \pi/2)/2),
\end{align}
which leads to the standard Ginsparg-Wilson chiral symmetries. A definition of $\alpha'$ and a discussion concerning its relationship to the $\alpha$ defined in the spectral decomposition of $D_d$ will be given below; although for the purposes of constructing the blocking any function of the eigenvalue which satisfies $\alpha = \phi_d$ at $\phi = 0$ and $\alpha = \phi_0$ at $\phi_0 = \pi/2$ (which ensure that the $\gamma_5$ matrices have the correct continuum limit) will suffice. Another option is to use
\begin{align}
B_1^{(\eta)} =& \left(\frac{2 \cos\phi_0}{\lambda_0} \right)^{(1+\eta)/2}R((1+\eta)(\phi_0 - \pi/2)/2+(1-\eta)(\alpha' - \pi/2)/2) \nonumber\\=& \left(\frac{2 \cos\phi_0}{\lambda_0} \right)^{(1+\eta)/2}R((1+\eta)(\phi_0 - \alpha')/2 + \alpha' - \pi/2)\nonumber\\
\overline{B}_1^{(\eta)} =& \left(\frac{2 \cos\phi_0}{\lambda_0} \right)^{(1-\eta)/2} R((1-\eta)(\phi_0 - \pi/2)/2-(1-\eta)(\alpha' - \pi/2)/2) \nonumber\\=&\left(\frac{2 \cos\phi_0}{\lambda_0} \right)^{(1-\eta)/2} R((1-\eta)(\phi_d - \alpha')/2) \nonumber\\
B_2^{(\eta)} =& \left( \cos\phi_d \cos\alpha \right)^{(1+\eta)/2}R((1+\eta)(\phi_d - \alpha)/2)\nonumber\\
\overline{B}_2^{(\eta)} =& \left(\cos\phi_d \cos\alpha \right)^{(1-\eta)/2}R((1-\eta)(\phi_d - \alpha)/2),
\end{align} 
and these blockings will be explored in this work. In principle, one can construct various groups of blockings by adding one factor of $(1-\eta)(\alpha' - \pi/2)/2$, partly to $B_1$ and partly to $\overline{B}_2$ to the blockings of equation (\ref{eq:naive}). 

The eigenvalues, $\lambda^D$, of $D_1$ are contained in one half of the overlap eigenvalue circle, $\Re \lambda^D \le 1$, while the eigenvalues of $D_2$ are contained within the other half of the circle. The two sets of eigenvectors meet at $\lambda^D = 1\pm i$ and between them complete the circle. 

\subsection{Chiral symmetry}\label{sec:4.1}
The chiral symmetry operators are constructed as
\begin{align}
\gamma_{L1}^{(\eta)} =& \overline{B}_1^{(\eta)} \gamma_5 (\overline{B}_1^{(\eta)})^{-1}&\gamma_{R1}^{(\eta)} =& ({B}_1^{(\eta)})^{-1} \gamma_5 B_1^{(\eta)}\nonumber\\ 
\gamma_{L2}^{(\eta)} =& \overline{B}_2^{(\eta)} \gamma_{Ld} (\overline{B}_2^{(\eta)})^{-1}&\gamma_{R2}^{(\eta)} =& ({B}_2^{(\eta)})^{-1} \gamma_{Rd} B_2^{(\eta)}.
\end{align}
These can be written as (again, for the non-zero eigenvalues)
\begin{align}
\gamma_{L}^{(\eta)} = & R((1-\eta)(\phi - \alpha)) \gamma_5\nonumber\\
\gamma_{R}^{(\eta)} = & \gamma_5 R((1+\eta)(\phi - \alpha) + 2\alpha - \pi),
\end{align}
where it is to be understood that when $\phi > \pi/4$ $\phi$ refers to $\phi_0$ and $\gamma_L$ refers to $\gamma_{L1}$, while for $\phi < \pi/4$ $\phi$ refers to $\phi_d$ and $\gamma_L$ refers to $\gamma_{L2}$. Similarly, we shall write $D$ is place of either $D_1$ or $D_2$ depending on the eigenvalue under consideration. $\alpha$ was originally defined for $0 < \alpha < \pi/4$ in terms of the eigenvalues of $D_d + 2$,
\begin{gather}
\tan\alpha = \frac{\lambda_d}{2},
\end{gather}
while
\begin{gather}
\tan{2\phi_d} = \frac{\lambda_d/2}{m(1-\lambda_d^2/4)},
\end{gather}
so that $\tan{2\alpha} = m/2 \tan(2\phi)$. We can use this new definition of $\alpha$, which is equivalent to the original definition for $\phi < \pi/4$, for all $0\le\phi\le\pi/2$, and choose $\alpha'$ to be this extended definition of $\alpha$. Thus we have,
\begin{align}
\gamma_{L}^{(\eta)} =& \left[\cos((1-\eta)(\phi - \alpha)) \gamma_5 +  \sin((1-\eta)(\phi - \alpha)) \sign(\gamma_5(D -D^{\dagger}))\right]\nonumber\\
\gamma_{R}^{(\eta)} = &\left[\cos((1+\eta)(\phi - \alpha) + 2\alpha - \pi) \gamma_5 -  \sin((1+\eta)(\phi - \alpha) + 2\alpha - \pi) \sign(\gamma_5(D -D^{\dagger}))\right] 
\end{align}
These operators are continuous functions of $\phi$ at the point where the eigenvalues meet on the Ginsparg-Wilson circle. At $\alpha = \phi = \pi/2$, corresponding to the zero modes, $\gamma_L = \gamma_R = \gamma_5$, while for $\alpha = \phi = 0$, corresponding to the zero mode doublers, $\gamma_L = - \gamma_R = \gamma_5$, and these operators, though originally constructed only for the non-zero modes, are well defined for the zero modes and their doublers at all $\eta$ (the original form of the Ginsparg-Wilson symmetry only reduced to the correct form at $\alpha = \phi = 0$ for $\eta$ odd integer).  It is straight forward to show that these $\gamma_5$-matrices satisfy the Ginsparg-Wilson equation for the appropriate Dirac operator, $D_1$ or $D_2$. The locality of these operators is discussed in \ref{app:locality}. Since $(\gamma_L^{(\eta)})^2 = (\gamma_R^{(\eta)})^2 = 1$, there are clearly no difficulties with singularities, and both functions are well defined for all eigenvalues of the Dirac operator. However, there are apparently numerous square roots in the definition, in the matrix sign function and the trigonometric functions of $\phi$ and $\alpha$, and these may cause a problem with locality if they generate branch cuts which touch the real axis of the Fourier transform of the operators. Using the standard Ginsparg Wilson relation, we can write that
\begin{align}
\sign(\gamma_5(D-D^{\dagger})) = \frac{\gamma_5(D-D^{\dagger})}{2\sin 2\phi}
\end{align}
and, from the definition of $\alpha$,
\begin{align}
\sin 2\alpha =& \frac{m \sin 2\phi}{\sqrt{m^2 \sin^2 2\phi + 4 \cos^2 2\phi}}\nonumber\\
\cos 2\alpha =& \frac{2 \cos 2\phi}{\sqrt{m^2 \sin^2 2\phi + 4 \cos^2 2\phi}},
\end{align}
with $\cos^2 \phi = D^{\dagger} D/4$ and $\sin^2 \phi  = 1-D^{\dagger}D/4$, where $D$ again represents $D_1$ or $D_2$ as appropriate. A potential non-locality exists if there are odd powers of $\sin \phi$ or $\cos \phi$ in the constructions of $\gamma_L$ or $\gamma_R$, because these involve square roots of functions of the Dirac operator which can be zero for certain momenta. The square root in the expression for $\sin(2\alpha)$ and $\cos (2\alpha)$ will not lead to any non-locality because the Fourier transform of $m^2 (1-D^{\dagger}D/4) + D^{\dagger}D = 0$ has no real momenta solutions for $0<m<2$. We can therefore write that
\begin{align}
\gamma_{L}^{(\eta)} =& \gamma_5 [\cos((1-\eta)\phi) \cos((1-\eta)\alpha) + \sin((1-\eta)\phi) \sin((1-\eta)\alpha)] +\nonumber\\ &
\frac{\gamma_5(D-D^{\dagger})}{2\sin 2\phi} [\sin((1-\eta)\phi) \cos((1-\eta)\alpha) - \cos((1-\eta)\phi) \sin((1-\eta)\alpha)].
\end{align}
For all odd integer $\eta$, the coefficient of $\gamma_5$ is an analytic function of $\cos^2 2\phi$ while the coefficient of $\sign(\gamma_5(D-D^{\dagger}))$ is $\sin 2\phi$ multiplied by an analytic function of $\cos^2 2\phi$. Therefore $\gamma_L$ is an analytic even function of $\cos(2\phi)$ and thus local. Similarly, it can be shown that $\gamma_{R}^{(\eta)}$ is local for all odd integer $\eta$.

\subsection{$\mathcal{CP}$ symmetry}\label{sec:4.2}
$\mathcal{CP}$ symmetry for these fermion fields will be modified for the blocked fields. The transformations of $\gamma_L$ and $\gamma_R$ under $\mathcal{CP}$ and the blockings $B$ and $\overline{B}$ can be constructed from the transformation laws of $\gamma_5$ and the Dirac operator,
\begin{align}
\gamma_R^{(\eta)} \rightarrow& - W(\gamma_5 \gamma_R^{(\eta)} \gamma_5)^TW^{-1}\nonumber\\
\gamma_L^{(\eta)} \rightarrow& - W(\gamma_5 \gamma_L^{(\eta)} \gamma_5)^TW^{-1}\label{eq:CP1}\\
B^{\eta} \rightarrow & W (B^{(\eta)})^T W^{-1}\nonumber\\
\overline{B}^{\eta} \rightarrow & W (\overline{B}^{(\eta)})^T W^{-1}.
\end{align}
$W$ is the continuum $\mathcal{CP}$ operator defined in equation (\ref{eq:cpforpsid}).
If
\begin{align}
\psi_1^{(\eta)} =& (B_1^{(\eta)})^{-1} \psi_0\nonumber\\
\overline{\psi}_1^{(\eta)} =&  \overline{\psi}_0(\overline{B}_1^{(\eta)})^{-1},
\end{align}
then, under a $\mathcal{CP}$ transformation,
\begin{align}
\psi_1^{(\eta)} \rightarrow & -(\overline{\psi}_1 G_1^{-1}\Gamma_1^{\dagger} )^T W^{-1}\nonumber\\
\overline{\psi}_1^{(\eta)} \rightarrow & W( G_1 \Gamma_1  {\psi}_1)^T. 
\end{align}
with
\begin{align}
\Gamma_1^{(\eta)} =& \left(\frac{2\cos\phi_0}{\lambda_0}\right)^{-\eta}(\overline{B}_1^{(\eta)})^{-1} {B}_1^{(\eta)} = R(\eta(\phi_0 - \alpha) + \alpha- \pi/2)\nonumber\\
G_1 =& \left(\frac{2\cos\phi_0}{\lambda_0}\right)^{\eta}
\end{align}
For $\psi_2$, taking note of modified $\mathcal{CP}$ transformation for $\psi_d$ (equation (\ref{eq:cpforpsid})), the transformation law is 
\begin{align}
\psi_2^{(\eta)} \rightarrow & -(\overline{\psi}_2G_2^{-1}\Gamma_2^{\dagger} )^T W^{-1}\nonumber\\
\overline{\psi}_2^{(\eta)} \rightarrow & W(G_2\Gamma_2  {\psi}_2)^T. \label{eq:CP2}
\end{align}
with
\begin{align}
\Gamma_2^{(\eta)} =& \left(\frac{2\cos\phi_d}{\sqrt{a^2\lambda_d^2 + 4}}\right)^{-\eta}\left(\overline{B}_2^{(\eta)}\right)^{-1}\Gamma_d {B}_2^{(\eta)} = R(\eta(\phi_d - \alpha) + \alpha - \pi/2)\nonumber\\
=& [\cos(\eta(\phi_d - \alpha) + \alpha - \pi/2) + \nonumber\\
&\phantom{lotsofspace}\sin(\eta(\phi_d - \alpha) + \alpha - \pi/2) \gamma_5 \sign(\gamma_5(D_2 - D_2^{\dagger}))](1-\phi_2\phi_2^{\dagger}) + \phi_2\phi_2^{\dagger}.\nonumber\\
G_2 =& \left(\frac{2\cos\phi_d}{\sqrt{a^2\lambda_d^2 + 4}}\right)^{\eta}
\end{align}
Once again, $[\Gamma_1^{(\eta)},D_1] = 0$, $[\Gamma_2^{(\eta)},D_2] = 0$ and
\begin{align}
(\Gamma_1^{(\eta)})^{\dagger} \gamma_5 \gamma_{R1}^{(\eta)} \Gamma_1^{(\eta)} =& \gamma_{L1}^{(\eta)}\nonumber\\
(\Gamma_2^{(\eta)})^{\dagger}\gamma_5 \gamma_{R2}^{(\eta)} \gamma_5 \Gamma_2^{(\eta)} =& \gamma_{L2}^{(\eta)}.\label{eq:CP4}
\end{align}
As with the previous action, $\Gamma_2$ is non-local; however this non-locality is the same as observed in the original theory. In the limit $a\rightarrow 0$, this transformation correctly reduces to the original $\mathcal{CP}$ transformation, which, as argued before, is all that is required. These two modified $\mathcal{CP}$ functions are of precisely the same form, and we can again write
\begin{align}
\Gamma^{(\eta)} =& [\cos(\eta(\phi - \alpha) + \alpha - \pi/2) + \nonumber\\
&\phantom{lotsofspace}\sin(\eta(\phi - \alpha) + \alpha - \pi/2) \gamma_5 \sign(\gamma_5(D - D^{\dagger}))](1-\phi_2\phi_2^{\dagger}) + \phi_2\phi_2^{\dagger}.
\end{align}
This function reduces in the continuum limit to $R(\alpha - \pi/2)$ for $\phi > \pi/4$ (corresponding to the alternative action for the physical fermion fields) and $\Gamma_d$ for the doublers. 

\subsection{Chiral gauge theories}\label{sec:4.3}
The overlap chiral gauge theory can be constructed by applying the blockings to the continuum chiral gauge theory
\begin{align}
\frac{1}{2}[\overline{\psi}_0 D_0 (1+\gamma_5) \psi_0 +& \overline{\psi}_d (D_d+2)(1+\gamma_{Rd}) \psi_d] \nonumber\\= &\frac{1}{2}[\overline{\psi}_1^{(\eta)}\overline{B}_1^{(\eta)} D_0 (1+\gamma_5) B_1^{(\eta)}\psi_1^{(\eta)} + \overline{\psi}_2^{(\eta)}\overline{B}_2^{(\eta)} (D_d+2)(1+\gamma_{Rd}) B_2^{(\eta)}\psi_2^{(\eta)}]\nonumber\\
=&\frac{1}{2}[\overline{\psi}_1^{(\eta)} D_1 (1+\gamma_{R1}^{(\eta)}) \psi_1^{(\eta)} + \overline{\psi}_2^{(\eta)} D_2(1+\gamma_{R2}^{(\eta)}) \psi_2^{(\eta)}].
\end{align}
Using equations (\ref{eq:CP1}), (\ref{eq:CP2}), and (\ref{eq:CP4}), it is straight forward to show that this action is invariant under $\mathcal{CP}$ symmetry:
\begin{align}
\frac{1}{2}[\overline{\psi}_1^{(\eta)} D_1 (1+\gamma_{R1}^{(\eta)})& \psi_1^{(\eta)} + \overline{\psi}^{(\eta)}_2 D_2(1+\gamma_{R2}^{(\eta)}) \psi_2^{(\eta)}]\nonumber\\\rightarrow &
\frac{1}{2}[\overline{\psi}^{(\eta)}_1(\Gamma_1^{(\eta)})^{\dagger} (1-\gamma_5\gamma_{R1}^{(\eta)}\gamma_5)D_1 \Gamma_1^{(\eta)} \psi^{(\eta)}_1 + \overline{\psi}^{(\eta)}_2(\Gamma_2^{(\eta)})^{\dagger} (1-\gamma_5\gamma_{R2}^{(\eta)}\gamma_5)D_2 \Gamma_2^{(\eta)}\psi_2^{(\eta)}]\nonumber\\
=&\frac{1}{2}[\overline{\psi}^{(\eta)}_1 (1-\gamma_{L1}^{(\eta)})D_1  \psi^{(\eta)}_1 + \overline{\psi}^{(\eta)}_2 (1+\gamma_{L2}^{(\eta)})D_2\psi_2^{(\eta)}]\nonumber\\
=&\frac{1}{2}[\overline{\psi}_1^{(\eta)} D_1 (1+\gamma_{R1}^{(\eta)}) \psi_1^{(\eta)} + \overline{\psi}^{(\eta)}_2 D_2(1+\gamma_{R2}^{(\eta)}) \psi_2^{(\eta)}]
\end{align}
Note in concluding this proof that the eigenvectors $\phi$ and $\phi_2$ change their chirality under $\mathcal{CP}$.

The demonstration that the measure is gauge and $\mathcal{CP}$ invariant follows the same method as the previous section. Again, we can construct the basis from the eigenvectors $g^{\pm}_{1}$ and $g^{\pm}_{2}$ of $\gamma_{R1}$ and $\gamma_{R2}$, using the relation $\gamma_L^{(\eta)} = = \Gamma_{\{1,2\}}^{\dagger}\gamma_5\gamma_R^{(\eta)}\gamma_5 \Gamma_{\{1,2\}}$ to write the eigenvectors of $\gamma_{L{1,2}}$ in terms of the eigenvectors of $\gamma_{R\{1,2\}}$, $\Gamma_{\{1,2\}}$ and $\gamma_5$, and expanding
\begin{align}
\frac{1}{2}\psi^{(\eta)}_2(1+\gamma_{R2}^{(\eta)}) =& \sum_i c_i g_{2i}^+ + \sum_i c_i^{(0)} g_{2i}\nonumber\\
\frac{1}{2}\psi^{(\eta)}_1(1+\gamma_{R1}^{(\eta)}) =& \sum_i c_i g_{1i}^+ + \sum_i c_i^{(0)} g_{0i}\nonumber\\
\frac{1}{2}\overline{\psi}^{(\eta)}_2(1-\gamma_{L2}^{(\eta)}) =& \sum_i c_i (g^-_{1i})^{\dagger}\Gamma_2^{\dagger}\gamma_5 + \sum_i c_i^{(0)} g_{2i}^{\dagger}\nonumber\\
\frac{1}{2}\overline{\psi}^{(\eta)}_1(1-\gamma_{L1}^{(\eta)}) =& \sum_i c_i (g^-_{-i})^{\dagger}\Gamma_1^{\dagger}\gamma_5 + \sum_i c_i^{(0)} g_{0i}^{\dagger}.
\end{align} 
Proof that this basis is $\mathcal{CP}$ invariant proceeds using an identical analysis to the previous section (with merely a substitution of the $\gamma_{\{L,R\}}$ and $\Gamma$ variables). Proof that the measure is gauge invariant is a little more involved, requiring an analysis of both the fermion and anti-fermion fields, but again the argument follows that of the previous section. 

\section{Lattice overlap fermions}\label{sec:5.1}
A `lattice' overlap operator (still constructed in the continuum, but expressed so that it is dominated by interactions between fermion fields on the lattice sites so that it is equivalent to the overlap operator constructed on the lattice) can be obtained by blocking the continuum overlap action. A decomposition of the continuum operator into lattice modes and off-lattice modes is made, and the kernel of the matrix sign function within the lattice operator constructed so that the off-lattice modes have infinite mass. After constructing a suitable kernel operator on the lattice with the right properties, RG blockings are found to map to and from the continuum Dirac operator. The kernel is controlled by a parameter $\zeta$ so that at $\zeta \rightarrow \infty$ it reduces to the Wilson lattice operator, while at all other $\zeta$ the Fourier transform of the Kernel and its inverse remain non-zero and finite (except for the necessary pole at $p=0$). This kernel is then put into the overlap formula, and the limit $\zeta \rightarrow \infty$ is taken to give the lattice overlap operator, $D$. It can then be shown that the Fourier transform of the Blockings required to construct this overlap operator from the continuum, and the inverses of the blockings, remain analytic~\cite{Cundy:2009ab}. We again stress the assumption, leaving the important discussion of this point to a future work, that the determinant of these blockings (which gives the change in the fermionic measure) can be absorbed into a renormalisation of the gauge fields and Yang Mills action.

$D$ can be written in matrix form as
\begin{gather}
D = 2 \cos\theta \left(\begin{array}{l l} \cos\theta & \sin\theta\\ -\sin\theta & \cos\theta\end{array}\right)
\end{gather}
with $\tan\theta = 2\sqrt{1-D^{\dagger}D/4}/{\sqrt{D^{\dagger}D}}$. 

I introduce a projection operator $P_d$, where
\begin{gather}
P_d = \frac{1}{2}\left(1 +\sign(D^{\dagger}D - 2)\right) \label{eq:Pd}.
\end{gather}
$[P_d,D] = [P_d,\gamma_5] = 0$. Note that this projection operator is non-local, due to the non-analyticity at $D^{\dagger}D = 2$ or $\theta = \pi/4$. However, since this operator will only be used as a tool during the construction of the action, and will not appear in the action or observables, this non-locality is unimportant. $P_d$ can be used to project the doublers from the physical modes in the lattice overlap operator: $P_d = 1$ for $\theta < \pi/4$ (the doublers\footnote{Defined as those eigenvector or momentum states where the slope of the dispersion relation for $D-D^{\dagger}$ has an opposite sign to the eigenvalue or momentum}) and 0 for $\theta > \pi/4$ (the physical modes). Also, $P_d^2 = P_d$ and $P_d(1-P_d) = 0$, so this operator can be used to project a fermion field onto a particular subspace. 

The continuum overlap action can be written in a matrix form as
\begin{gather}
\begin{array}{l l}(\overline{\psi}_1&\overline{\psi}_2)\\
&\end{array}
\left(\begin{array}{l l} D_1&0\\0&D_2\end{array}\right)\left(\begin{array}{l} \psi_1\\ \psi_2\end{array}\right),
\end{gather}
and the blockings from the continuum overlap Dirac operators $D_1$ and $D_2$ to the lattice overlap can be written as
\begin{align}
\left(\begin{array}{l} \psi_1\\ \psi_2\end{array}\right) =& \left(\begin{array}{l l} Z&0\\0&Z_d\end{array}\right)\left(\begin{array}{l} (1-P_d)\psi\\ P_d\psi\end{array}\right)\nonumber\\
\begin{array}{l l} (\overline{\psi}_1 &\overline{\psi}_2)\\ \end{array} =& 
\begin{array}{l l} (\overline{\psi} (1-P_d) &\overline{\psi} P_d)\\ \end{array} \left(\begin{array}{l l} Z^{\dagger}&0\\0&Z_d^{\dagger}\end{array}\right),
\end{align}   
where $Z$ and $Z_d$ are unitary operators mapping eigenvectors of $D_1$ and $D_2$ to eigenvectors of $D$~\cite{Cundy:2009ab}. We have to choose which eigenvectors of $D_1$ and $D_2$ to map onto the lattice and which to project to infinite mass in the lattice formulation. The simplest mapping connects eigenvectors whose eigenvalues are at the same point around the Ginsparg-Wilson circle; thus eigenvectors of $D_1^{\dagger}D_1$ whose eigenvalues are parametrised by $\phi_0$ are mapped onto eigenvectors of $D^{\dagger}D$ with $\theta \leftarrow \phi_0$. Similarly, for the doublers, $\theta \leftarrow \phi_d$. Again, we write that $\tan(2\alpha) = m/2\tan (2\theta)$, and obtain blockings that transform the continuum action based on Dirac operators $D_0$ and $D_d+2$ to the overlap operator $D$.
\begin{align}
B^{(\eta)} =&\left(\begin{array}{l l} ZB_a^{(\eta)}&0\\0 & Z_dB_b^{(\eta)}\end{array}\right)\nonumber\\
\overline{B}^{(\eta)} =& \left(\begin{array}{l l}\overline{B}_a^{(\eta)}Z^{\dagger} &0\\0&  \overline{B}_b^{(\eta)}Z_d^{\dagger}\end{array}\right)
\end{align}
with
\begin{align}
&\left.\begin{array}{l}
B_a^{(\eta)} = \cos((1+\eta)(\theta - \alpha)/2 + \alpha - \pi/2) +\\ \phantom{lots of space space lots}\gamma_5 \sign(\gamma_5(D-D^{\dagger})) \sin((1+\eta)(\theta - \alpha)/2 + \alpha - \pi/2)\\
\overline{B}_a^{(\eta)} = \cos((1-\eta)(\theta - \alpha)/2) + \gamma_5 \sign(\gamma_5(D-D^{\dagger})) \sin((1-\eta)(\theta - \alpha)/2)\end{array}\right\} \theta > \pi/4
\nonumber\\
& \left. \begin{array}{l}B_b^{(\eta)} = \cos((1+\eta)(\theta - \alpha)/2) + \gamma_5 \sign(\gamma_5(D-D^{\dagger})) \sin((1+\eta)(\theta - \alpha)/2)\\
\overline{B}_b^{(\eta)} = \cos((1-\eta)(\theta - \alpha)/2) + \gamma_5 \sign(\gamma_5(D-D^{\dagger})) \sin((1-\eta)(\theta - \alpha)/2)
\end{array} \right\}\theta < \pi/4
\end{align}
The zero modes of the lattice overlap Dirac operator are at $\theta = \pi/2, P_d = 0, \alpha = \pi/2$, and their doublers at $\theta = 0, P_d = 1, \alpha = 0$. Therefore these blockings (and their inverses, which are straight forward to construct) are well defined at the zeros of $\gamma_5(D-D^{\dagger})$ because the coefficient multiplying $\sign(\gamma_5(D-D^{\dagger}))$ is zero.  
\subsection{Chiral symmetry}\label{sec:5.2}
From these blockings, and the chiral symmetry for the continuum overlap operator we can construct the associated chiral symmetry operators
\begin{align}
\gamma_R^{(\eta)} = (1-P_d)&\left(\begin{array}{l l} \cos((1+\eta)(\theta-\alpha) + 2\alpha - \pi)&\sin((1+\eta)(\theta-\alpha)+2\alpha - \pi)\\
\sin((1+\eta)(\theta-\alpha) + 2\alpha - \pi)&-\cos((1+\eta)(\theta - \alpha) + 2\alpha - \pi)\end{array}\right)
+\nonumber\\
&P_d\left(\begin{array}{l l} \cos((1+\eta)(\theta - \alpha)+2\alpha-\pi)&\sin((1+\eta)(\theta-\alpha)+2\alpha-\pi)\\
\sin((1+\eta)(\theta - \alpha )+2\alpha - \pi)&-\cos((1+\eta)(\theta - \alpha) + 2\alpha - \pi)\end{array}\right)
\nonumber\\
\gamma_L^{(\eta)} = (1-P_d)&\left(\begin{array}{l l} \cos((1-\eta)(\theta-\alpha))&-\sin((1-\eta)(\theta-\alpha))\\
-\sin((1-\eta)(\theta-\alpha))&-\cos((1-\eta)(\theta - \alpha))\end{array}\right)
+\nonumber\\
&P_d\left(\begin{array}{l l} \cos((1-\eta)(\theta - \alpha))&-\sin((1-\eta)(\theta - \alpha))\\
-\sin((1-\eta)(\theta - \alpha))&-\cos((1-\eta)(\theta-\alpha))\end{array}\right)
\end{align}
Or,
\begin{align}
\gamma_R^{(\eta)} =&  \gamma_5R((1+\eta)(\theta-\alpha) + 2\alpha - \pi)\nonumber\\
= & \gamma_5 \cos((1+\eta)(\theta-\alpha) + 2\alpha - \pi) + \sign(\gamma_5(D-D^{\dagger})) \sin((1+\eta)(\theta-\alpha) + 2\alpha - \pi)\nonumber\\
\gamma_L^{(\eta)} = &R((1-\eta)(\theta-\alpha))\gamma_5\nonumber\\
=&\gamma_5\cos((1-\eta)(\theta - \alpha)) - \sign(\gamma_5(D-D^{\dagger})\sin((1-\eta)(\theta - \alpha)).
\end{align}
These chiral symmetry operators are continuous and well defined for all eigenvalues of the Dirac operator, as at $\theta = 0$ or $\pi/2$ the coefficient of $\sign(\gamma_5(D-D^{\dagger}))$ is zero, while at $\theta = \pi/4$ the coefficient of $P_d$ and thus $\sign(D^{\dagger}D-2)$ is zero. In each of these cases, the ambiguity in the definition of the matrix sign function is unimportant. Following a similar argument to the previous section, it can be demonstrated that these operators are local for all odd integer $\eta$.

Using the matrix formulation, it is easy to show that the Ginsparg-Wilson equation
\begin{gather}
\gamma_L^{(\eta)} D + D \gamma_R^{(\eta)} = 0\nonumber
\end{gather}  
is satisfied for the overlap operator, and therefore these operators can be used to define a group of lattice chiral symmetries.

It is useful to express these operators at odd integer in a more practical formulation. Given that 
\begin{gather}
\gamma_L^{(\eta)}
= \left(\begin{array} {l l} 
\cos((1-\eta)(\theta - \alpha)) & \sin((1-\eta)(\theta - \alpha))\\
-\sin((1-\eta)(\theta - \alpha))&\cos((1-\eta)(\theta - \alpha))
\end{array}\right)\gamma_5,
\end{gather}
we have
\begin{gather}
\gamma_L^{(\eta_1)}\gamma_L^{(\eta_2)} = \gamma_L^{(\eta_1 - \eta_2 + 1)} \gamma_5.
\end{gather}
selecting $\eta_2 = -1$, we have
\begin{gather}
\gamma_L^{(\eta_1 + 2)} = \gamma_L^{(\eta_1)} \gamma_L^{(-1)} \gamma_5.
\end{gather}
A direct calculation gives
\begin{align}
\gamma_L^{(-1)}\gamma_5 =& \frac{(2-m)(D-1)(D^{\dagger}D/2 - 1) + m}{\sqrt{(m^2-4)(1-D^{\dagger}D/4) D^{\dagger} D + 4}},\nonumber\\
\gamma_L^{(1)} =& \gamma_5
\end{align}
and the construction of the complete set of $\gamma_L^{(\eta)}$ for integer $\eta$ follows. $\gamma_R^{(\eta)}$ can then be constructed from the Ginsparg-Wilson relation and the commutator $[D,D^{\dagger}]=0$:
\begin{gather}
\gamma_R^{(\eta)} = -D^{-1} \gamma_L^{(\eta)} D = -\gamma_L^{(\eta)} \frac{D}{D^{\dagger}} = \gamma_L^{(\eta)}(1-D). 
\end{gather}
It is clear that this group of chiral symmetries, valid for any $0<m<2$ (regardless of the mass that was originally used within the sign function of the kernel) is distinct from the group described in the introduction. For example, at those eigenvectors where $D^{\dagger}D = 2$, this $\gamma_L^{-1}\gamma_5 = 1$ while for the first group $\gamma_L^{-1}\gamma_5$ was $(1-D)$. There is also a third group of chiral symmetries, which can be constructed using the same process from an additional continuum chiral symmetry
\begin{align}
\gamma'_{Rd} = &\gamma_5\nonumber\\
\gamma'_{Ld} = &\gamma_5\left(1-\frac{4}{D_d + 2}\right).
\end{align}
This would result in swapping the definitions of $\gamma_L^{(\eta)}$ and $\gamma_5 \gamma_R^{(\eta)} \gamma_5$.

\subsection{$\mathcal{CP}$ symmetry}\label{sec:5.3}
The $\mathcal{CP}$ symmetry for the fermion fields associated with these blockings can be constructed easily using the matrix notation. As the procedure mirrors exactly the discussion of the previous section, we will only outline the main results. The fermion fields transform as
\begin{align}
\psi^{(\eta)} \rightarrow& -W(\overline{\psi}^{(\eta)}G^{-1}\Gamma^{\dagger})^T\nonumber\\
\overline{\psi}^{(\eta)}\rightarrow & (G\Gamma \psi^{(\eta)})^T W^{-1}
\end{align}
with, 
\begin{align}
&\Gamma^{(\eta)} = [\cos(\eta(\theta - \alpha) + \alpha - \pi/2) + \nonumber\\
&\phantom{lotsandlotsofspace}\sin(\eta(\theta - \alpha) + \alpha - \pi/2) \gamma_5 \sign(\gamma_5(D - D^{\dagger}))](1-\phi_2\phi_2^{\dagger}) + \phi_2\phi_2^{\dagger}.\nonumber\\
&G =  \left(Z_0^{\dagger}\frac{1}{\sqrt{D_0^{\dagger} D_0}}Z_0 \cos\theta \frac{1}{2}(1- \sign(\cos(2\theta)) + Z_d^{\dagger}\frac{1}{\sqrt{D_d^{\dagger}D_d+4}}Z_d \cos\theta \frac{1}{2}(1+ \sign(\cos(2\theta))\right)^{\eta}
\end{align}
$W$ remains the continuum $\mathcal{CP}$ operator defined in equation (\ref{eq:cpforpsid}).
Again, this is well defined at the potential discontinuities, $\theta = \pi/2$. At $\theta = 0$, $\Gamma$ is non-local; but this non-locality is derived from the non-locality in the target continuum theory. The operator mapping between the lattice operator and the continuum operator is local, and $\Gamma^{(\eta)}$ does not appear in any actions or observables; so this non-locality has no effect on any observables nor prevent the continuum limit from being well defined and is thus unimportant. 

Clearly 
\begin{gather}
[\Gamma,D] = 0\label{eq:CPL1}
\end {gather}
 as they are both SO(2) matrices. Equally, the $\mathcal{CP}$ transformations of the $\gamma_{L,R}$ operators are
\begin{align}
\gamma_R^{(\eta)} \rightarrow &  -(\Gamma^{(\eta)})^{\dagger} \gamma_L^{(\eta)}\Gamma^{(\eta)}\nonumber\\
\gamma_L^{(\eta)} \rightarrow &  -(\Gamma^{(\eta)})^{\dagger} \gamma_R^{(\eta)}\Gamma^{(\eta)}.\label{eq:CPL2}
\end{align}

\subsection{Chiral gauge theories}\label{sec:5.4}
The chiral gauge theory can be constructed in the same way by blocking the continuum fermion fields. The resulting action is
\begin{gather}
S = \overline{\psi}^{(\eta)} D (1+\gamma_R^{(\eta)}) \psi^{(\eta)}.
\end{gather} 
Using equations (\ref{eq:CPL1}) and (\ref{eq:CPL2}) it is manifest that this action is gauge invariant.

The measure again can be constructed from the eigenvectors of $\gamma_{R}^{(\eta)}$:
\begin{align}
\frac{1}{2}\psi^{(\eta)}(1+\gamma_{R}^{(\eta)}) =& \sum_i c_i g^+_{i} + \sum_i c_i^{(0)} g_{0i}\nonumber\\
\frac{1}{2}\overline{\psi}^{(\eta)}(1-\gamma_{L}^{(\eta)}) =& \sum_i c_i (g^-_{i})^{\dagger}\Gamma^{\dagger}\gamma_5 + \sum_i \overline{c}_i^{(0)} g_{0i}^{\dagger}.
\end{align} 
Proof that the measure is $\mathcal{CP}$ and gauge invariant again follows the method of the previous sections.

\section{Conclusions}
I have constructed another class of Ginsparg-Wilson chiral symmetry operators, for lattice overlap fermions, this time related to a continuum theory containing a physical fermion and (this is the novelty) a doubler fermion with a mass of the order of the cut-off. Construction of lattice $\mathcal{CP}$ symmetries and chiral gauge theories follows. This continuum action provides a more natural fit with the lattice action.

I have not here discussed the physical implications of the multitude of lattice chiral symmetries, whether the original group discovered by Mandula, or the new group presented in this work. Our hope is that this work will, however, form the basis of such a discussion, which we hope to address in a future work.

\section*{Acknowledgements}
N. Cundy is supported by the BK21 program  of the NRF grant
funded by the Korean government (MEST).
The research of W.~Lee is supported by the Creative Research
Initiatives program (3348-20090015) of the NRF grant funded by the
Korean government (MEST).
\appendix
\section{Locality}\label{app:locality}
\subsection{The Paley-Wiener theorem}

If we consider a function $F(x)$ in the continuum, then we can construct the Fourier transform $\tilde{F}(p)$ according to
\begin{align}
\tilde{F}(p) =& \int_{-\infty}^{\infty} d^4x F(x) e^{i(p,x)}\nonumber\\
F(x) = & \frac{1}{(2\pi)^4} \int_{-\infty}^{\infty} d^4p \tilde{F}(p) e^{-i(p,x)}.
\end{align}
If $F$ possesses an O(4) rotational symmetry, then it suffices to align one of the axis in the direction of $x$, for example we can write that $p_t$ is parallel to $x$.
\begin{align}
F(x) = & \frac{1}{(2\pi)^4} \int_{-\infty}^{\infty} d^4p \tilde{F}(p_t,p_x,p_y,p_z) e^{-ip_tx_t}.
\end{align}
For $x>0$, the integral can be completed in the complex plane around the lower half circle as long as $\tilde{F}(p)$ is finite on this circle and zero at $p = \pm \infty$. If $\tilde{F}(p)$ is analytic along this contour, which, most importantly, includes being analytic along the real axis, then the integral over $p$ is given by the sum over the residues of $\tilde{F}(p)$ and the integral around the branch cuts in the lower half complex plane. If there are no branch cuts,  and if the poles are at $\pi_i = \pi_i^{re} \pm i \pi_i^{im}$ ($\pi_i^{re},\pi_i^{im}$ both real and $\pi_i^{im} > 0$; $\pi_i$ will in general be a function of the other components of the momenta), then $F(x)$ will have the form,
\begin{gather}
F(x) = \sum_i \int d^3 p \alpha_i(\pi_i,p_x,p_y,p_z) e^{-\pi_i^{im}|x|}.\label{eq:pw1}
\end{gather}     
Assuming that when the remaining integral is calculated or estimated (for example, by using the method of steepest descent) it retains the exponential form (which will certainly occur if $\pi_i^{im}$ is constrained to be greater than some positive number $\beta$), then
\begin{gather}
|F(x)| < \alpha e^{-\beta |x|}\label{eq:pw2}
\end{gather}
 for some positive $\alpha$ and $\beta$ and $F$ is local. If there are additionally branch cuts as well as poles, from momentum $\pi_{j\;\text{start}}$ (with imaginary component closest to zero) to $\pi_{j\;\text{end}}$ then the functional form of equation (\ref{eq:pw1}) will no longer be valid, and instead we must use
\begin{align}
F(x) = \int d\theta_1 d\theta_2 d\theta_3 &\sin^2{\theta_1}\sin{\theta_2} \bigg[\sum_i \alpha_i(\pi_i,p_x,p_y,p_z) e^{-\pi_i^{im}|x_t|} +\nonumber\\ 
& \sum_j \int_{\pi_{j\;\text{end}}}^{\pi_{j\;\text{start}}} d\pi \alpha^j(\pi,p_x,p_y,p_z) e^{-i\pi x_t}\bigg].
\end{align}
Since, at each point along the integral, $ |\alpha^j(\pi,p_x,p_y,p_z) e^{-i\pi x_t}| < C' e^{-p_{j\;\text{start}} |x_t|}$ for some positive $C'$, the whole integral is smaller than $C e^{-\pi_{j\;\text{start}}|x_t|}$ and the equation (\ref{eq:pw2}) still holds as long as the branch cut does not cross the real axis. Thus for any function in the continuum where the Fourier transform is analytic along the real axis, the function itself is at least exponentially local. This is, of course, the Paley-Wiener theorem~\cite{Paley-Wiener}. 

On the lattice, the momentum is bounded, $|p_{\mu}| < \pi/a$, so before the Paley-Weiner theorem can be applied it is necessary to transform  to a new momentum variable, $\hat{p}$ which is not bounded, for example using
\begin{gather}
  \frac{a}{2}\hat{p}_{\mu} = \tan\left(\frac{a}{2}{p}_{\mu}\right). 
\end{gather}
This gives
\begin{gather}
d p_{\mu} = d \hat{p}_{\mu} \frac{1}{1 + \left(\frac{a \hat{p}_{\mu}}{2}\right)^2}.\label{eq:latticepole}
\end{gather}
After this transformation of variables, the Paley-Wiener theorem can be derived in the same way; once again the result is that if $\tilde{F}(p)$ as analytic along the real axis for $-\pi/a<\hat{p}<\pi/a$, then the resulting operator $F(x)$ will be exponentially local or better. Equation (\ref{eq:latticepole}) indicates that any lattice operator for which this construction is valid will only be exponentially local, with a rate of decay inversely proportional to the lattice spacing, and cannot be ultra-local.

\subsection{Application}
Suppose that there is some function, which contains the matrix sign function, of a lattice Hermitian Dirac operator $K$,
\begin{gather}
F(K) = f(K) \sign(K),
\end{gather}
where $f$ is a local function (whose Fourier transform is analytic for real momentum), and we wish to determine whether or not $F(K)$ is local. we proceed by writing the matrix sign function in its integral form,
\begin{gather}
\sign(K) = \frac{1}{\pi} \int_{-\infty}^{\infty} dt\frac{K}{t^2 + K^2}.
\end{gather}
Next, just considering the case where $x>0$, we take the Fourier transform, twice.
\begin{gather}
{F}(K(0,x)) = \frac{1}{\pi(2\pi)^4} \int_{-\infty}^{\infty} dtd^4 \hat{p} \prod_{\mu}\frac{1}{1+ (a\hat{p}_{\mu}/2)^2} f(K(\hat{p})) \frac{K(\hat{p})}{t^2 + K(\hat{p})^2} e^{i2x \arctan (\hat{p}_{t} a/2)/a}.   
\end{gather} 
$K$ will contain some Dirac structure and gauge links, so before proceeding one will have to diagonalise this spinor structure and consider each eigenvalue separately. Here, we can write $K(\hat{p}) = \gamma_5a^{(a)}T^a (\hat{p}) + i b_{\mu}^{(a)} T^a(\hat{p}) \gamma_5 \gamma_{\mu} + c_{\mu\nu}^{(a)}T^a(\hat{p}) i\gamma_5 \sigma_{\mu\nu}$, with $a$, $b$ and $c$ real functions of the momentum and where $T^a$ are the Hermitian generators of the gauge group. This particular form is forced because $K$ is Hermitian and $\gamma_5 K$ invariant under $\mathcal{CP}$. It is then possible to diagonalise this operator, and consider each of the eigenvalues separately. We shall assume that this has been done. 

First we perform the integration over $\hat{p}_t$, using a contour integration. We can close the integrand over either the upper or lower hemisphere in the complex plane; for the upper hemisphere the contour integration will pick out those poles with imaginary part greater or equal to zero. The poles will in general be at some complex momenta $\tilde{p}_t^{(i)}(\hat{p}_y,\hat{p}_z,\hat{p}_x,t)$, and if the poles in $\hat{p}_t$, as a function of $t$ and the other components of the momentum, have non-zero imaginary part for all real $t$ and spacial momenta, then the function will be local. This is the case for any poles coming from $f(k)$, since this function is analytic for real momenta, and $1/(1+(ap/2)^2)$. The only poles which may have zero imaginary part are the solutions to $0 =  t^2 + K(\hat{p}_t,\hat{p}_x,\hat{p}_y,\hat{p}_z)^2$, and here only for $t=0$ since $K^2\ge 0$ for real momenta. Performing the integral over $p_t$ gives (where the sum is over those poles where $\Im(\tilde{p}_t^{(i)}) \le 0$)
\begin{align}
{F}(K(0,x)) = \frac{2 i}{(2\pi)^4}\sum_i \int &dt \int dp_x dp_y dp_z  f(K(\tilde{p}_0^{(i)})) \left(\left(\frac{\partial K}{\partial p_t}\right)_{p_t = \tilde{p}_t^{(i)}}\right)^{-1} \nonumber\\
&e^{-2ix \arctan(\tilde{p}_t^{(i)} a/2)/a}\prod_{\mu = x,y,z}\frac{1}{1+ (a\hat{p}_{\mu}/2)^2} \frac{1}{1+ (a\tilde{p}_{t}^{(i)}/2)^2}.   
\end{align}
I note that, if $\tilde{p}_t^{(i)} = \tilde{p}_{tr}^{(i)} + i\tilde{p}_{ti}^{(i)}$,  
\begin{align}
2ix \arctan(\tilde{p}_t^{(i)} a/2)/a =& (x/2a) \log\left(\frac{(1- a\tilde{p}_{ti}^{(i)}/2)^2 + (a\tilde{p}_{tr}^{(i)}/2)^2}{(1+ a\tilde{p}_{ti}^{(i)}/2)^2 + (a\tilde{p}_{tr}^{(i)}/2)^2}\right) +\nonumber\\
&\phantom{space} i(x/a) \left(\arctan\frac{a\tilde{p}_{tr}^{(i)}/2}{1-a\tilde{p}_{ti}^{(i)}/2} + \arctan\frac{a\tilde{p}_{tr}^{(i)}/2}{1+a\tilde{p}_{ti}^{(i)}/2}\right)
\end{align}
The integrals over $t$ and the spacial momenta can be performed at large $x$ by a saddle point approximation. Considering the integral over $t$ as an example, it will be dominated by those values of $t$ where
\begin{gather}
\Delta = (x/2a) \log\left(\frac{(1- a\tilde{p}_{ti}^{(i)}/2)^2 + (a\tilde{p}_{tr}^{(i)}/2)^2}{(1- a\tilde{p}_{ti}^{(i)}/2)^2 + (a\tilde{p}_{tr}^{(i)}/2)^2}\right) + \log(f(K(\tilde{p}_t^{(i)},p_x,p_y,p_z))) - \log (1+ (a \tilde{p}_t^{(i)}/2)^2)
\nonumber
\end{gather}
is maximised (this expression is negative) given the restriction that $p_{ti}^{(i)} > 0$. Unless either of the second two terms becomes infinite, at large $x$ the expression will become dominated by the first term, and the question becomes finding the maxima of the coefficient of $x$, and seeing if any of these maxima are at zero. 
\begin{gather}
\log\left(\frac{(1- a\tilde{p}_{ti}^{(i)}/2)^2 + (a\tilde{p}_{tr}^{(i)}/2)^2}{(1+ a\tilde{p}_{ti}^{(i)}/2)^2 + (a\tilde{p}_{tr}^{(i)}/2)^2}\right) = 0
\end{gather} 
is only solved at $\tilde{p}_{ti}^{(i)} = 0$ or $\tilde{p}_{tr}^{(i)} = \infty$ while $\tilde{p}_{ti}^{(i)}$ remains finite and non-zero or $\tilde{p}_{ti}^{(i)} = \infty$. If either $p_{tr}$ or $p_{ti} = \infty$ then the real part of $\log (1+ (a \tilde{p}_t^{(i)}/2)^2)= \infty$, and $\Delta$ is certainly not maximised. The only option remaining is whether there is a solution with $\tilde{p}_{ti}^{(i)} = 0$, which requires that there is a real solution to $K^2(p) + t^2 = 0$. At these points $\partial \Delta/\partial t$ need not be 0, but all the other maxima of $\Delta$ may be found by solving   $\partial \Delta/\partial t = 0$. If there is a real momenta solution to $K(\hat{p}) = 0$, then the solution to $K^2(\hat{p}) +t^2 = 0$ will have a solution $\tilde{p}_{ti}^{(i)} \propto t$. Otherwise, $\tilde{p}_{ti}^{(i)} = a +  O(t)$ for some $a>0$, and the function is local.  

Suppose that there is a real solution to $K(\hat{p}) = 0$. Let us further suppose that around $t = 0$, either $f(K(\tilde{p}_t(t,p_x,p_y,p_z),p_x,p_y,p_z))$ or any one of its derivatives with respect to $t$ is non-zero (this excludes the trivial case of $f=0$). The integral over $t$ at large $x$ will then be the sum over several terms; exponential decays over $x$ with rates determined by the minima of $\arctan(\tilde{p}_t^{(i)} a/2)/a$ and the contribution from around $t=0$, which will be given by an integral of the form
\begin{gather}
\int_{-\infty}^{\infty} dt (a+b't + \ldots) e^{-x|t|\alpha'/a},
\end{gather}
for some $a$, $b'$ and $\alpha'$. This will decay according to $a/x$, not good enough for the function to be considered local in the proper sense of the word, even though it has a sharp decay (which, in a numerical simulation on a finite volume lattice, may be hidden behind the exponential decays and the finite arithmetical precision). It is not enough for the function $F(K)$ (if it is a member of the class of functions we are interested in) to be well defined: for $F$  to be local, it must either contain no square root of $K^2$ or $K^2 = 0$ should have no real solutions.

The free Wilson operator, used in the construction of the overlap operator, $K = \gamma_5 (i \gamma_{\mu} \sin(p_{\mu}) + 2\sum_{\mu} \sin^2(p_{\mu}/2) - m)$ has eigenvalues $\sqrt{\sin^2 p_{\mu} + (2\sum_{\mu} \sin^2(p_{\mu}/2) - m))^2}$ which cannot be zero, so $K^2 = 0$ has no real solutions. This will also hold for the interacting Wilson operator on a smooth enough gauge field. However, this is not true for the matrix sign functions and projectors used in the construction of the $\gamma$-operators $\sign(D-D^{\dagger})$ and $\sign (D^{\dagger}D - 2)$. In both these cases, on particular gauge fields, there will be real momentum solutions to $D-D^{\dagger} = 0$ and $D^{\dagger}D =2$. 

Additionally, we shall require to know whether functions such as $\sqrt{c \pm \sqrt{d}}$ are local, where $c$ and $d$ are some functions of $D^{\dagger} D$ which are analytic for real momenta. The procedure is the same as before. If $c^2 > d$, we can write that
\begin{gather}
\sqrt{c \pm \sqrt{d}} = \frac{1}{\pi}\int_{-\infty}^{\infty} dt \left(\frac{t^2 + c}{(t^2 + c)^2 - d} \mp \frac{1}{\pi}\int_{-\infty}^{\infty}dt' \frac{d}{((t^2 + c)^2- d)((t')^2 + d)} \right), 
\end{gather}
and the same reasoning shows that the function will be local if there are no real momenta solutions to $d$ and $c^2 - d$ are greater than some positive value. Additionally, if $c > 0$ and $d > 0$ and $d < \Lambda$ for some finite positive $\Lambda$ then it is possible to write that $\sqrt{d} = R(d)/P(d)$ where $R$ and $P$ are positive convergent polynomials of $D^{\dagger} D$ (for example, found using Zolotarev's expansion). Then
\begin{gather}
\sqrt{c + \sqrt{d}} = \sqrt{cP + R} P^{-1/2},
\end{gather}
which is local as both $cP + R$ and $P$ are greater than some positive number.

For odd integer $\eta$, as discussed in section \ref{sec:4.1}, the $\gamma_5$ matrices and mapping of $\mathcal{CP}$ symmetry from the lattice to the continuum are local. For these values of $\eta$, we can write that
\begin{align}
\gamma_L^{(\eta)} = &\frac{(2-m)(D-1)(D^{\dagger}D/2 - 1) + m}{\sqrt{(m^2-4)(1-D^{\dagger}D/4) D^{\dagger} D + 4}}
\nonumber\\
\gamma_R^{(\eta)} = & \gamma_L^{(\eta)} (1-D).
\end{align} 
Since this is a polynomial in the overlap operator and $ m^2(1-D^{\dagger}(p)D(p)/4) + D^{\dagger}(p) D(p) = 0$ has no real solutions given that $0 <D^{\dagger}D < 4$, is is clear that these functions are local. Similarly, the $\mathcal{CP}$ matrix generated during the blocking between overlap fermions and continuum overlap fermions is $\Gamma = Z_0^{\dagger} D_1^{(-\eta)} Z_0 D^{(\eta)} (1-\sign(\cos(2\theta))/2 + Z_d^{\dagger} D_2^{(-\eta)} Z_d D^{(\eta)} (1+\sign(\cos(2\theta))/2$ which is local for integer $\eta$ as the zeros of $D^{\eta}$ and the poles of $Z_0^{\dagger} D_1^{-\eta} Z_0$ are at the same momentum and the same order in $p$ and $Z_0^{\dagger} D_1^{-\eta} Z_0  (1-\sign(\cos(2\theta))/2 + Z_d^{\dagger} D_2^{-\eta} Z_d (1+\sign(\cos(2\theta))/2$ is a continuous function of the momentum. 

The matrix mapping between the $\mathcal{CP}$ transformation of overlap continuum fermions and continuum fermions is, for example, 
\begin{align}
G_1^{(\eta)}\Gamma_1^{(\eta)} =& \left(\frac{\cos\phi}{\lambda_0}\right)^{\eta} R(\eta(\phi - \alpha) + \alpha - \pi/2)(1-\phi_2\phi_2^{\dagger}) + \phi_2\phi_2^{\dagger}.\nonumber\\
=& \left(\frac{\cos\phi}{\lambda_0}\right)^{\eta}R(\eta(\phi - \alpha)) R(\alpha - \pi/2)(1-\phi_2\phi_2^{\dagger}) + \phi_2\phi_2^{\dagger}.
\end{align} 
The $\mathcal{CP}$ matrix in the continuum can be written as (equations (\ref{eq:Gammacont1}) and (\ref{eq:Gammacont1}))
\begin{gather}
R(\alpha - \pi/2)(1-\phi_2\phi_2^{\dagger}) + \phi_2\phi_2^{\dagger}.
\end{gather}
We wish to map from the lattice action to the continuum action $\overline{\psi}'_0 D_0 \psi'_0 + \overline{\psi}_d(D_d + 2) \psi_d$. We require 
\begin{enumerate}
\item
 the mapping between the $\gamma_5$ matrices and $\mathcal{CP}$ matrices is local. This follows from the locality of the operators. The $\mathcal{CP}$ matrix is non-local on both the continuum and the lattice; to demonstrate that the mapping between the lattice $\mathcal{CP}$ matrix and continuum $\mathcal{CP}$ matrix is straight-forward, but requires to be established.
\item The construction of this continuum action is valid. It is sufficient to demonstrate that $\gamma_{L0}$, $\gamma_{R0}$, $\Gamma_0$ and the blockings are local (i.e. exponentially local with the rate of decay proportional to the cut-off). 
\end{enumerate}
To demonstrate the locality of $R(\eta(\theta - \alpha)$ (which is the mapping between the continuum and lattice $\mathcal{CP}$ matrices), it is useful to construct a Dirac operator 
\begin{align}
D_{\alpha} = 2\cos \alpha R(\alpha) = \cos(2\alpha) + 1 + \frac{\sin(2\alpha)}{2 \sin (2\phi)} (D_1-D_1^{\dagger}).
\end{align}
Using the definition of $\alpha$ on the lattice
\begin{align}
\cos(2\alpha) = & \frac{2\cos(2\theta)}{\sqrt{(m^2 -4)\sin^2 (2\theta) + 4}}
\nonumber\\
\sin(2\alpha) = &\frac{m\sin(2\theta)}{\sqrt{(m^2 -4)\sin^2 (2\theta) + 4}} 
\nonumber\\
\cos (\alpha) = &\frac{1}{\sqrt{2}}\left(1+ \frac{D_1^{\dagger}D_1 - 2}{\sqrt{4 + (m^2 - 4)D_1^{\dagger}D_1 (1-D_1^{\dagger}D_1/4)}}\right)^{1/2} 
\end{align}
and, by using the argument of this section, it can be shown that $D_{\alpha}$ is local. Furthermore, $D_{\alpha}$ only contains poles at $\alpha = \pi/2$ and the eigenvectors corresponding to these poles are the same as the zero modes of $D_1$. Therefore $(D_1/D_{\alpha})^\eta$ is local for any integer $\eta$. 
Since $\cos(\phi) = \sqrt{D^{\dagger}D}/2$ and $\cos(\alpha)$ does not have any zeros except at $D^{\dagger}D = 0$, $(\cos(\theta)/\cos(\alpha))^{\eta}$ is local for all integer $\eta$. Since $R(\eta(\theta - \alpha)) = ((\cos \alpha D)/(\cos\theta D_{\alpha}))^{\eta}$, this function is local.

Now we consider the continuum action $\overline{\psi}' D_0 \psi$, and whether the blockings used to construct these fermion fields are local. In the continuum,
\begin{align}
\cos(2\alpha) = & \frac{2\cos(2\phi)}{\sqrt{(m^2 -4)\sin^2 (2\phi) + 4}}
\nonumber\\
\sin(2\alpha) = &\frac{m\sin(2\phi)}{\sqrt{(m^2 -4)\sin^2 (2\phi) + 4}}
\end{align}
where, for $\phi > \pi/4$,
\begin{align}
\sin(2\phi) =& \frac{\sqrt{D_0 D_0^{\dagger}}}{\sqrt{m^2 (1-D_0 D_0^{\dagger}/4)^2 + D_0 D_0^{\dagger}}}\nonumber\\
\cos(2\phi) =& -\frac{m(1-D_0 D_0^{\dagger}/4)}{\sqrt{m^2 (1-D_0 D_0^{\dagger}/4)^2 + D_0 D_0^{\dagger}}}.
\end{align}
This gives
\begin{align}
\cos(\alpha - \pi/2) = & \frac{2}{\sqrt{4+D_0D_0^{\dagger}}}\nonumber\\
\sin(\alpha - \pi/2) = & -\frac{\sqrt{D_0 D_0^{\dagger}}}{\sqrt{4+D_0D_0^{\dagger}}}\nonumber\\
R(\alpha - \pi/2) = & \frac{2}{\sqrt{4+D_0D_0^{\dagger}}}(1-D_0)\\
\cos(\alpha/2 - \pi/4) = & \frac{1}{\sqrt{2}} \left(1+ \frac{2}{\sqrt{4+D_0 D_0^{\dagger}}} \right)^{1/2}\nonumber\\
\sin(\alpha/2 - \pi/4) = & \frac{1}{\sqrt{2}} \left(1- \frac{2}{\sqrt{4+D_0 D_0^{\dagger}}} \right)^{1/2}\nonumber\\
R(\alpha/2 - \pi/4) = & \frac{1}{\sqrt{2}} \left(1+ \frac{2}{\sqrt{4+D_0 D_0^{\dagger}}} \right)^{1/2} + \frac{1}{\sqrt{2}} \left(1- \frac{2}{\sqrt{4+D_0 D_0^{\dagger}}} \right)^{1/2} \frac{D_0}{\sqrt{D_0^{\dagger} D_0}}\nonumber\\
=&\frac{1}{\sqrt{2}} \left(1+ \frac{2}{\sqrt{4+D_0 D_0^{\dagger}}} \right)^{1/2} + \frac{1}{\sqrt{2}} \left(1+ \frac{2}{\sqrt{4+D_0 D_0^{\dagger}}} \right)^{-1/2} D_0 
\end{align}
And, using the results of this section, it can be seen that the blockings used to generate $\overline{\psi}_0'D_0 \psi_0'$, $R(\alpha/2 - \pi/4)$ are local, as are the $\mathcal{CP}$ and $\gamma_5$ matrices which are proportional to $R(\alpha - \pi/2)$. This statement only applies to the physical continuum fields; it does not hold for the transformations of the continuum doubler fermions. Thus the continuum action used as a basis for this group of Ginsparg-Wilson symmetries.

\bibliographystyle{elsarticle-num}
\bibliography{weyl}

\end{document}